\documentclass[aps,floats,prl,superscriptaddress,
twocolumn]{revtex4}

\usepackage{amsfonts,amsmath,mathrsfs}
\usepackage{amssymb} \usepackage{dsfont}
\usepackage{bm} \usepackage{bbm}
\usepackage{dcolumn}
\usepackage{epsfig} \usepackage{latexsym}
\usepackage{float}

\begin{document}

\title{Concentration-of-measure theory for structures and fluctuations of waves}

\author{Ping Fang}
\affiliation{Institute for Advanced Study, Tsinghua University, Beijing 100084,China}
\affiliation{CAS Key Laboratory of Theoretical Physics and Institute of Theoretical Physics,
Chinese Academy of Sciences, Beijing 100190, China}

\author{Liyi Zhao}
\affiliation{Institute for Advanced Study, Tsinghua University, Beijing 100084,China}

\author{Chushun Tian}
\email{ct@itp.ac.cn}
\affiliation{CAS Key Laboratory of Theoretical Physics and Institute of Theoretical Physics,
Chinese Academy of Sciences, Beijing 100190, China}

\date{\today}

\begin{abstract}

The emergence of nonequilibrium phenomena in individual complex wave systems has long been of fundamental interests. Its analytic studies remain notoriously difficult. Using the mathematical tool of the {\it concentration of measure} (CM), we develop a theory for structures and fluctuations of waves in individual disordered media. We find that, for both diffusive and localized waves, fluctuations associated with the change in incoming waves (``wave-to-wave'' fluctuations) exhibit a new kind of universalities, which does not exist in conventional mesoscopic fluctuations associated with the change in disorder realizations (``sample-to-sample'' fluctuations), and originate from the coherence between the natural channels of waves -- the {\it transmission eigenchannels}. Using the results obtained for wave-to-wave fluctuations, we find the criterion for almost all stationary scattering states to exhibit the same spatial structure such as the diffusive steady state. We further show that the expectations of observables at stationary scattering states are independent of incoming waves and given by their averages with respect to eigenchannels. This suggests the possibility of extending the studies of thermalization of closed systems to open systems, which provides new perspectives for the emergence of nonequilibrium statistical phenomena.

\end{abstract}

\maketitle

Recent studies on the foundations of equilibrium statistical mechanics \cite{Lebowitz06,Popescu06,Deutsch91,Srednicki94,Rigol08,Rigol16,Borhonovi16} have shed new light on the long-standing problem of nonequilibrium phenomena in individual (quantum) wave systems, where neither fictitious ensembles nor reservoirs exist \cite{von Neumann29}. Many of them \cite{Deutsch91,Srednicki94} rely on a conjecture of Berry, i.e., in closed systems random scattering can render waves structureless on large spatial scales \cite{Berry77}. This wave property gives rise to a basic feature of thermal equilibrium phenomena in individual closed systems, i.e., spatial homogeneity. Yet, a major topic of nonequilibrium statistical mechanics is concerned with various spatial structures in open systems \cite{Reichl,Dorfman,Haenggi05,Lepri03,Gaspard96}. This sharp contrast motivates exploring in depth spatial structures of waves, i.e., scattering states, in individual open systems, which may also pave a way for extending the studies of the relations between spatial and entanglement structures in an ensemble of open disordered systems -- a new aspect of the fundamentals of nonequilibrium statistical mechanics \cite{Huse18} -- to an individual member.

In fact, spatial structures and fluctuations of waves in open disordered media are central topics of mesoscopic physics \cite{Rossum99,Sheng95,Mirlin00,Efetov97,Akkermans}.
However, most theoretical efforts have been focused on disorder ensembles; the significance of waves in individual disordered media has been emphasized only recently \cite{Dogariu09,Genack13}. The common wisdom of using self-averaging or the ergodic hypothesis to connect certain properties of individual disordered media to their disorder averages \cite{Hanggi98,Lifshits88} essentially requires the thermodynamic limit, and cannot be applied to study fluctuations in mesoscopic scales. It remains a challenge to construct a theory for wave statistics in individual mesoscopic systems, where rich fluctuation phenomena of the wave origin can be driven, e.g., by changing the incoming wave. Such wave-to-wave fluctuations differ from well understood sample-to-sample fluctuations \cite{Altshuler91,Rossum99,Sheng95,Beenakker97,Hanggi98,Lifshits88,Altshuler85,Lee85,Mirlin00,Kamenev,Efetov97,Akkermans}. Their in-depth studies are of both fundamental and practical importance. Indeed, in individual mesoscopic systems fluctuations and irreversibility have been known to be closely related \cite{Altshuler96}. In addition, the strength of wave-to-wave fluctuations determines whether a generic wave can represent the behaviors of most waves, i.e., is typical: in recent years the arising of {\it typicality} has appeared as a central topic in statistical mechanics \cite{von Neumann29,Lebowitz06,Popescu06}, but been restricted to closed systems so far. On the other hand, wave statistics in individual open disordered media has found many optical applications \cite{Gigan10,Fink12,Choi12,Rotter17}.

Recently, the CM \cite{Milman86,Ledoux01,Boucheron13} -- ``one of the great ideas of analysis in our time'' \cite{Talagrand96} -- has been adopted to study statistical phenomena in individual closed classical \cite{Bievre17} and quantum \cite{Popescu06,Winter06,Guarneri15} systems. The CM is rooted in high-dimensional geometry. The idea can be illustrated by the unit sphere, for which the area of the sphere becomes more and more concentrated around the equator as the dimension increases. Eventually in high dimensions the entire area almost concentrates around the equator. This property can then be visualized by real-valued functions over the sphere with nice continuity properties, through their concentration around some constant value. When the sphere is replaced by a general high-dimensional geometric body (e.g., the Euclidean space) and the area measure by others (e.g., the Gaussian measure), similar results follow. This idea opens new perspectives of probability theory \cite{Talagrand96,Boucheron13,Ledoux01}. It allows us not only to study variables with complicated dependence on random variables instead of being their sum, but also to obtain results which are nonasymptotic, i.e., do not require the limit of large number of variables. A detailed introduction of CM is given in section S0 of supplemental materials (SM) \cite{SM}.

\begin{figure}
\includegraphics[width=8.7cm] {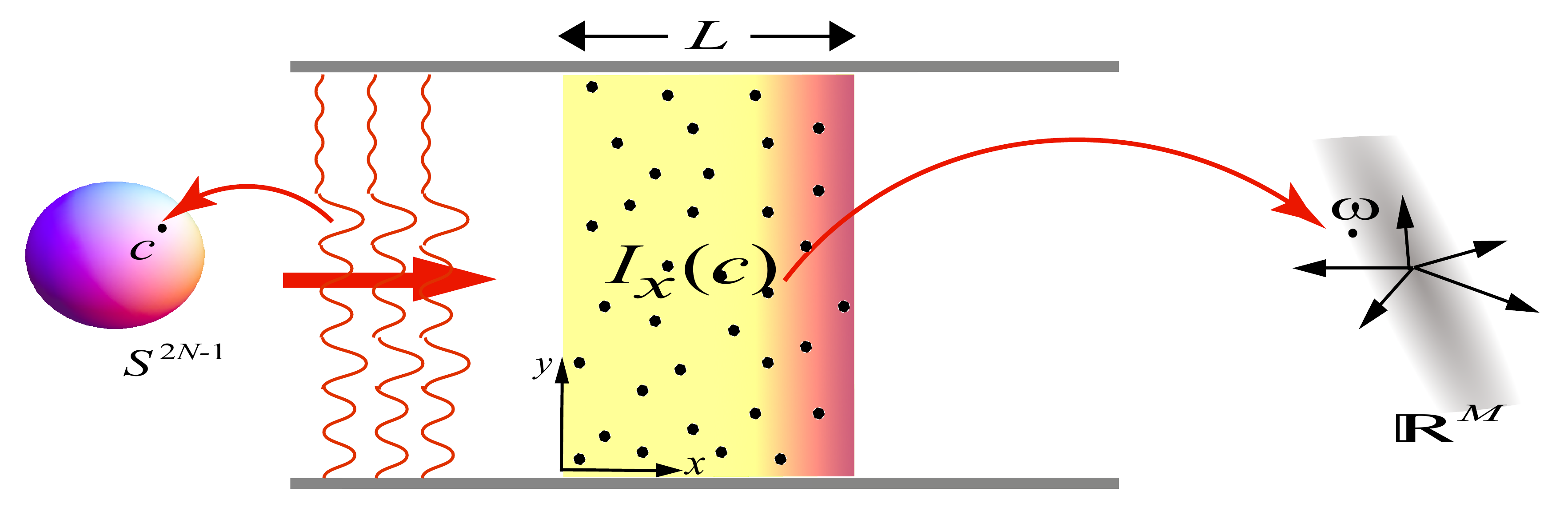}
\caption{A wave is launched into a disordered medium of $N$ channels. In this setting there are two geometric bodies that provide the basis for applying CM, namely, the sphere $S^{2N-1}$ constituted by distinct incoming current amplitudes $c$ and the Euclidean space $\mathds{R}^M$ by distinct disorder realizations $\omega$.}
\label{fig:2}
\end{figure}

\begin{figure}[h]
\includegraphics[width=8.7cm] {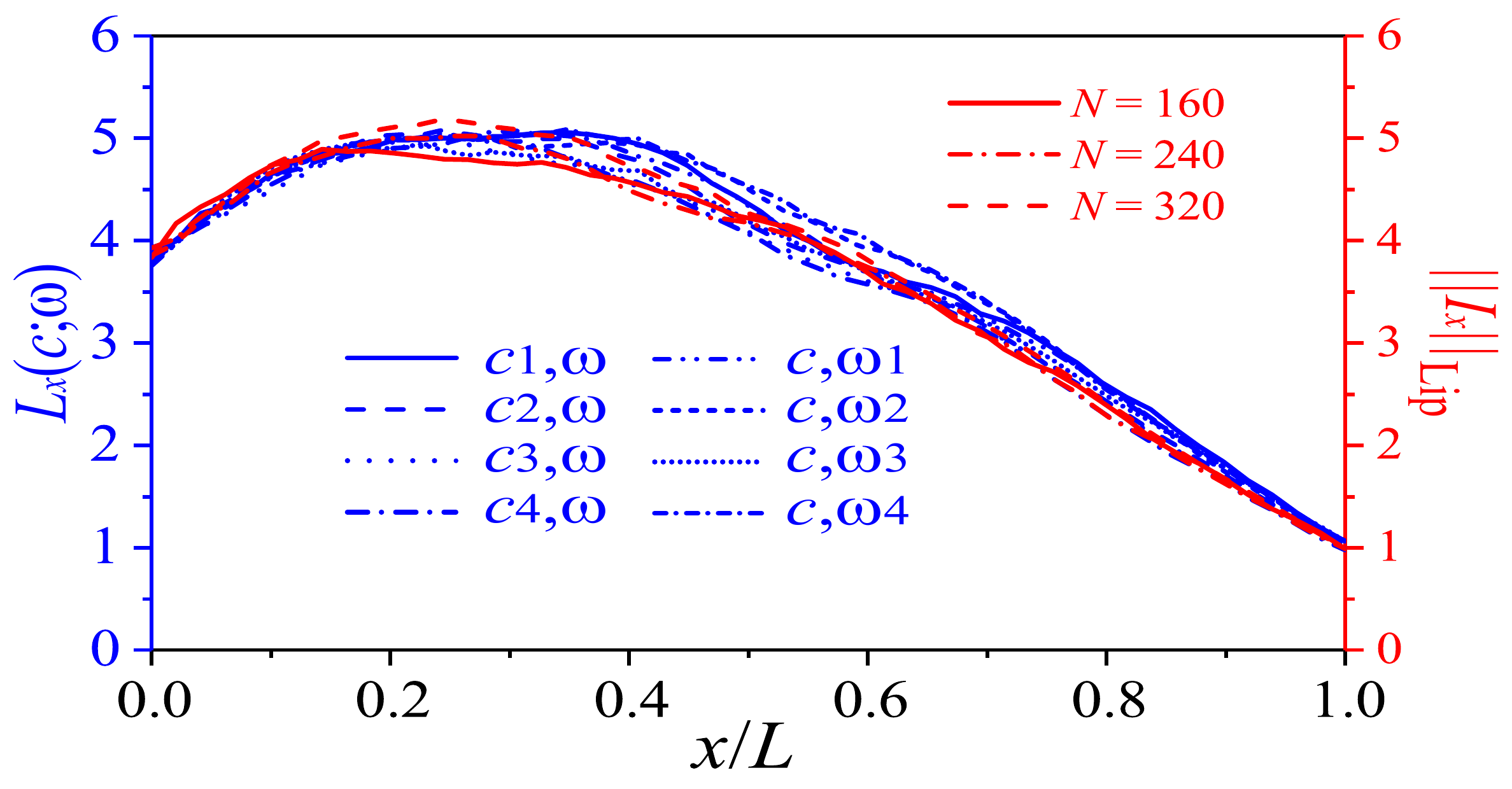}
\caption{Using Eqs.~(\ref{eq:4}) and (\ref{eq:3}), we calculate $L_x(c;\omega)$ at $N=800$ for $4$ randomly chosen $c$ at fixed $\omega$ and for $4$ randomly chosen $\omega$ at fixed $c$, respectively, and calculate $\|I_x\|_{\rm Lip}$ for $3$ large $N$. All profiles collapse into the a single curve. $L=50$}
\label{fig:4}
\end{figure}

In this work we employ CM to explore universal statistical phenomena of waves in individual open disordered media. We launch a classical wave of circular frequency $\Omega$ and carrying unit energy flux into a finite medium with $N$ ($=$$\frac{\Omega}{\pi}$$\times$the width) channels and length $L$ \cite{note3,note5} (Fig.~\ref{fig:2}). Keeping the disorder realization fixed, but allowing the incoming wave to vary gives rise to various wave-to-wave fluctuations. Below, most attentions are paid to the fluctuations of the spatial structure of scattering state corresponding to the incoming current amplitude $c$, i.e., the depth ($x$) profile $I_x(c)$ of energy density integrated over the cross section. We develop a CM theory of wave-to-wave fluctuations. Physically, it provides information on a single stationary scattering state in a single disordered medium and the differences of behaviors between a disorder ensemble and an individual member; technically, instead of traditional impurity diagrams \cite{Rossum99,Sheng95,Akkermans} and field theories \cite{Mirlin00,Kamenev,Efetov97}, its key components are various {\it concentration inequalities} \cite{Boucheron13} of observables [e.g., $I_x(c)$]. Armed with the developed theory we achieve the following results:
\begin{itemize}
  \item We find that compared to conventional sample-to-sample fluctuations, wave-to-wave fluctuations exhibit a number of ``anomalies''. In particular, irrespective of regimes of wave propagation (diffusive, localized, etc.), the distribution of $I_x(c)$ is always {\it sub-Gaussian}, i.e., has an (upper) tail decaying at least as fast as a Gaussian tail [Eq.~(\ref{eq:1})]. Contrary to this, for sample-to-sample fluctuations of observables such as total transmission, as waves are more and more localized the distribution tail decays slower and slower, and the shape of the tail changes dramatically \cite{Mirlin00,Genack97}.
  \item Furthermore, we find that the wave-to-wave fluctuations of $I_x(c)$ are governed by an $x$-dependent curve $\|I_x\|_{\rm Lip}$, which arises from the phase coherence between distinct eigenchannels -- the natural channels for wave propagation in disordered media \cite{Choi11,Tian15}. In contrast, the sample-to-sample fluctuations of $I_x(c)$ ($c$ fixed) are governed by the conductance \cite{Rossum95} known to equal the number of open eigenchannels \cite{Imry86}. For diffusive waves we find that the curve is universal with respect to disorder realizations $\omega$ at large $N$ (cf.~Fig.~\ref{fig:4}).
  \item We find the criterion (\ref{eq:7}) for almost all stationary scattering states to exhibit the same spatial structure, i.e., a nonequilibrium steady state, and show that it can be readily satisfied for diffusive waves.
  \item We show that the expectations of generic observables at stationary scattering states are independent of incoming waves and given by their averages with respect to eigenchannels [Eq.~(\ref{eq:13})], and find the corresponding criterion.
\end{itemize}
\noindent The results summarized above are attributed to general wave properties, and thus apply to both classical and quantum waves. (This is similar to Anderson localization applying to both classical and quantum waves \cite{Mirlin00,Sheng95}.)

While we are not aware of any studies of the applications of CM to wave propagation and scattering in disordered media, we begin with a general discussion on how the high-dimensional geometry emerges from the present setting (Fig.~\ref{fig:2}), and further provides a basis for applying CM. For simplicity we consider a two-dimensional ($2$D) medium.
A disordered dielectric configuration $\delta\epsilon (x,y)$ is embedded into the air background. So the wave field $E(x,y)$ satisfies the Helmholtz equation \cite{note5},
\begin{eqnarray}\label{eq:15}
    \left\{\partial_x^2 +\partial_y^2 +\Omega^2 [1+
    \delta\epsilon (x,y)]\right\}E(x,y)=0.
\end{eqnarray}
Given $N$ channel bases, an incoming current amplitude is a projection represented by $N$ complex coefficients: $(c_1$,$c_2$,...,$c_N)$$\equiv$$c$. As the incoming wave carries a unit energy flux, we have $\sum_{n=1}^N$$|$$c_n$$|^2$$=$$1$. So all $c$ constitute a high-dimensional geometric body, the unit sphere $S^{2N-1}$. Next, we discretize the medium into a lattice of $M$ points. The $M$ values $\{\omega_{(x,y)}$$\equiv$$-$$\Omega^2$$\delta$$\epsilon$$($$x$,$y$)$\}$ then constitute the coordinate of another high-dimensional geometric body, the Euclidean space $\mathds{R}^M$, a point in which corresponds to a disorder realization $\omega$$\equiv$$\{$$\omega_{(x,y)}$$\}$.
As observables depending on $c$ (respectively $\omega$) define real-valued functions over $S^{2N-1}$ (respectively $\mathds{R}^M$),
we can apply CM to them and probe the properties of the wave-to-wave (respectively sample-to-sample) fluctuations by corresponding observables.

{\it Construction of the theory.} We divide the construction into five steps so that the readers may keep track of it. In each step, we outline the derivations and present the key results; motivations and (or) physical implications are also discussed. Technical details and expanded discussions are relegated to the self-contained SM.

{\it Step 1 -- formulation of the problem.} Below we choose the eigenchannels as the basis. Thus we first introduce the eigenchannel briefly, and this is done in three steps as follows \cite{Tian15}. (i) Consider the transmission matrix $t\equiv \{t_{ab}\}$, which transforms an incoming current amplitude $c$ into a transmitted current amplitude given by $tc$. Both current amplitude are vectors in the space spanned by the ideal waveguide modes $\varphi_{a}(y)$, with $a$ the mode index. The matrix element $t_{ab}=-i\sqrt{\tilde v_a\tilde v_b} \langle
x=\infty a|G
|x'=-\infty b\rangle
$ \cite{note6}, where $G$ is the retarded Green's function and $\tilde v_a$ is the group velocity of mode $a$. By the singular value decomposition, $t=\sum_{n=1}^N u_n \sqrt{\tau_n} v_n^\dagger$, we obtain a transmission eigenvalue spectrum $\{\tau_n\}$ ($\tau_n$ decreases with $n$.) and two mutual orthogonal unit vectors
$\{u_n\}
$ and $\{v_n\}
$.
(ii) Replacing $x=\infty$ in $t_{ab}$ above by arbitrary $x\in [0,L]$, we make the extension: $t\rightarrow t(x)\equiv \{t_{ab}(x)\}$. This gives the vector field inside the medium, $E_{n}(x)
=t(x)v_n$, excited by $v_n$. (iii) The triple: $(\tau_n,v_n, E_n(x)\equiv\{E_{na}(x)\})$ defines an eigenchannel, which are completely fixed by $\omega$ and $\Omega$ \cite{note7}. Each channel transmits waves with transmission coefficient $\tau_n$, and has a specific $2$D spatial structure, namely, the energy density profile $|E_{n}(x,y)|^2$ with $E_n(x,y)=\sum_{a=1}^N E_{na}(x)\varphi_a^*(y)$. Integrating out $y$ we reduce $|E_{n}(x,y)|^2$ to a one-dimensional ($1$D) structure,
\begin{equation}\label{eq:16}
\begin{array}{c}
  W_{\tau_n}(x)\equiv \int dy|E_{n}(x,y)|^2=E_{n}^\dagger(x)\cdot E_{n}(x),
\end{array}
\end{equation}
where $\cdot$ refers to a scalar product.

To proceed we introduce the precise definition of $I_x(c)$ \cite{note2}: $I_x(c)\equiv \int dy E^*(x,y)\hat v_xE(x,y)$, where $\hat v_x\equiv (1-\partial_{\Omega y}^2)^{\frac{1}{2}}$ is a scalar (not vector) operator accounting for the absolute value of the group velocity in waveguide modes. Treating $\omega_{(x,y)}$ as a scattering potential, we apply the scattering theory of waves \cite{Newton82} to Eq.~(\ref{eq:15}) and find $\hat v_x^{\frac{1}{2}}E(x,y)=\sum_{a=1}^N(t(x)c)_a\varphi^*_a(y)$. Here $c=\sum_{n=1}^N c_n v_n$ is the incoming current amplitude in the eigenchannel representation. Thus
\begin{eqnarray}\label{eq:10}
\begin{array}{c}
  I_x(c)=\sum_{n,n'=1}^N c_n^*c_{n'} E_{n}^\dagger(x)\cdot E_{n'}(x).
\end{array}
\end{eqnarray}
This defines a family of real-valued functions over $S^{2N-1}$, and $x$ labels these functions. Note that at $x<L$ the vectors $E_{n}(x)$ are not orthogonal.

Then the problem is: {\it For fixed $\omega$, does $I_x(c)$ exhibit universal behaviors when $c$ varies}? A natural idea is to calculate all the cumulants of $I_x(c)$ and find the distribution. But, one then needs to calculate an infinite number of products of $E_{n}^\dagger\cdot E_{n'}$, and sum up their contributions, which is a formidable task especially for small $N$. The CM allows a different route, which we follow below.

{\it Step 2 -- Lipschitz continuity: a building block of CM.} This is the concept that formalizes the ``nice continuity properties'' of real-valued functions mentioned in the introductory part. Let a generic space $\mathscr{C}$ be equipped with the Euclidean metric $\|\cdot\|$. For $f$$:$$\mathscr{C}$$\rightarrow$$\mathds{R}$, if
\begin{eqnarray}
\begin{array}{c}
  \|f\|_{{\rm Lip}}\equiv{\rm sup}_{z,z'
  }\frac{|f(z)-f(z')|}{\|z-z'\|}<\infty\\
  \Leftrightarrow \,|f(z)-f(z')|\leq \|f\|_{{\rm Lip}}\,\|z-z'\|,
\end{array}
\label{eq:11}
\end{eqnarray}
where `${\rm sup}$' stands for the least upper bound, then $f(z)$ is said to have the Lipschitz continuity or be Lipschitz, and $\|f\|_{{\rm Lip}}$ is called the Lipschitz constant. As we will see below, even though $f$ has a very complicated dependence on $c$, its wave-to-wave fluctuations are controlled by a single parameter, i.e., $\|f\|_{{\rm Lip}}$.

{\it Step 3 -- Concentration inequality for $I_x(c)$ and results for general waves.} With the preparations above we are now ready to introduce the following result of CM:\\
\\
\noindent L$\acute{\rm e}$vy's lemma \cite{Winter06,Milman86}. {\it Let $\mu$ be the uniform probability measure over $S^{2N-1}$,
and $f: S^{2N-1}\rightarrow \mathds{R}$ be Lipschitz. Then the probability for the
deviation between $f$ and its mean $\int f d\mu$ to exceed $\varepsilon$ is $\leq 2e^{-
\frac{\delta\varepsilon^2 N}{\|f\|_{{\rm Lip}}^2}}$,
where $\delta$ is some positive absolute constant.}\\

\begin{figure}
\includegraphics[width=8.7cm] {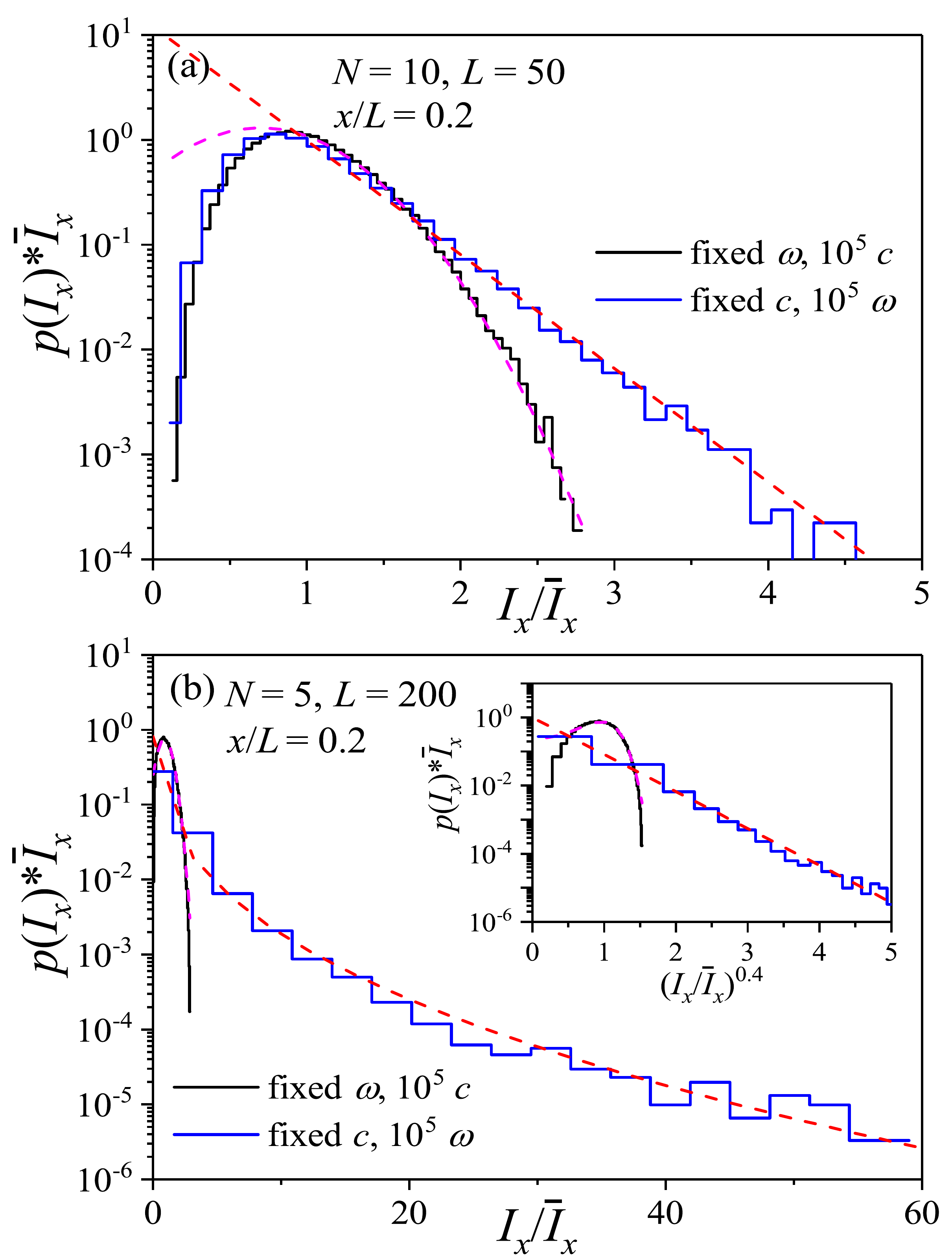}
\caption{Quasi $1$D simulations show that, for both diffusive (a) and localized (b) waves, the distribution of wave-to-wave fluctuations of $I_x(c)$ (black histograms) displays a tail well fit by a Gaussian distribution (pink dashed lines),
while that of sample-to-sample fluctuations (blue histograms) is exponential (a) and stretched-exponential with a stretching exponent of $0.4$ (b), respectively (red dashed lines).
In (b) the main panel and inset differ in the horizontal axis. The localization length $N\ell$ \cite{Dorokhov84,Mello88} is $130$ in (a) and $65$ in (b). ${\bar I}_x$$=$$\int$$I_xp$$($$I_x$$)$$d$$I_x$.}
\label{fig:3}
\end{figure}

\noindent This means that $f$ concentrates around $\int fd\mu$ with a rate increasing rapidly with $\frac{N}{\|f\|_{{\rm Lip}}^2}$. We stress that $\|f\|_{{\rm Lip}}$ depends on $N$ generally. By the lemma the distribution of $f$ is sub-Gaussian. But, unlike the central limit theorem, the lemma does not require the large $N$ limit, i.e., is nonasymptotic; instead, $N$ can be very small (cf.~Fig.~\ref{fig:3}).

To apply L$\acute{\rm e}$vy's lemma to $I_x(c)$, in SM (S1) we use Eq.~(\ref{eq:10}) to derive an analytic expression of the Lipschitz constant $\|I_x\|_{{\rm Lip}}$ of $I_x(c)$. The result reads
\begin{eqnarray}
\begin{array}{c}
  \|I_x\|_{{\rm Lip}}={\rm sup}_c L_x(c;\omega),\\
  L_x=\pi \left(\sum_{n=1}^N|\sum_{n'=1}^N \!c_{n'}^* E_{n'}^\dagger(x)\cdot E_{n}(x)|^2-I_x^2\right)^{1/2},
\end{array}
\label{eq:4}
\end{eqnarray}
where $L_x$ depends on $c,\omega$ in general. According to Eq.~(\ref{eq:4}) $\|I_x\|_{{\rm Lip}}<\infty$. Combined with L$\acute{\rm e}$vy's lemma this gives the following concentration inequality,
\begin{equation}\label{eq:1}
\begin{array}{c}
  {\rm Pr}\left(|I_x(c)-W(x;\omega)|>\varepsilon\right)\leq 2e^{-
    \frac{\delta\varepsilon^2 N}{\|I_x\|_{{\rm Lip}}^2}}.
\end{array}
\end{equation}
Here $W(x;\omega)\equiv \int I_x(c)d\mu$ and ``Pr'' stands for probability. After simple algebra we reduce $W(x;\omega)$ to
\begin{eqnarray}
\begin{array}{c}
  W(x;\omega)=\frac{1}{N}\sum_{n=1}^N W_{\tau_n}(x).
\end{array}
\label{eq:12}
\end{eqnarray}
The results (\ref{eq:4})-(\ref{eq:12}) hold for general $N, \omega$, regardless of regimes of wave propagation (diffusive, localized, etc.).

From the inequality (\ref{eq:1}) we see that provided
\begin{equation}\label{eq:7}
    W(x;\omega)\gg \|I_x\|_{{\rm Lip}}/\sqrt{N},
\end{equation}
the wave-to-wave fluctuations of $I_x(c)$ are negligible and almost all incoming waves behave in essentially the same way: their energies are stored in distinct channels with an equal weight of $1/N$ and a nonequilibrium steady state universal with respect to $c$ namely $W(x;\omega)$ results, i.e., $I_x$$($$c$$)$$\approx$$W$$($$x$$;$$\omega$$)$. The state does not carry information on the phase coherence between eigenchannels. The phase coherence, as shown by the sub-Gaussian tail in (\ref{eq:1}) and Eq.~(\ref{eq:4}), enters into $\|I_x\|_{\rm Lip}$ and influences strongly wave-to-wave fluctuations (see {\it Step 5} for further discussions).

From the inequality (\ref{eq:1}) we also see that the distribution tail of $I_x(c)$ decays
at least as fast as a Gaussian tail. In contrast, in sample-to-sample fluctuations ($c$ fixed)
the distribution tail decays much slower, known (for $x=L$) to be exponential
for diffusive waves \cite{Rossum95} and log-normal for deeply localized waves \cite{Beenakker97,Mirlin00}.

{\it Step 4 -- Diffusive steady state and its fluctuations.} Below we use the general results (\ref{eq:4})-(\ref{eq:12}) to explore in-depth diffusive waves.
Numerical studies have shown that in quasi $1$D media \cite{note8} the disorder average of Eq.~(\ref{eq:12}) gives a diffusive steady state \cite{Tian15}, but it is difficult to (dis)prove analytically that without the averaging, this remains true for general geometry. Indeed, for large $N$ ($L$ and $\Omega$ fixed) the medium is a slab \cite{note8} and thus high-dimensional, but in high dimension the explicit form of $W_{\tau_n}(x)$, even its disorder average, is unknown; for small $N$, i.e., a short quasi $1$D medium \cite{note8}, the impacts of the sample-to-sample fluctuations on $W_{\tau_n}(x)$ have not yet been studied.

To study Eq.~(\ref{eq:12}) we start from large $N$ and establish a concentration inequality of $W(x;\omega)$. To this end we show in SM (S2) that, even for a single disordered slab, distinct eigenchannel structures $W_{\tau_n}(x)$ are described by a single formula \cite{Tian15}, that depends smoothly on the eigenvalue $\tau_n$ and was derived originally for an ensemble of quasi $1$D disordered media. Using this fact,
we show in SM (S3) that $W(x;\omega)$, which is a real-valued function over $\mathds{R}^M\ni\omega$, is Lipschitz, i.e.,
\begin{eqnarray}\label{eq:2}
  |W(x;\omega)-W(x;\omega')|\leq
  \tilde c(x)N^{-\frac{1}{2}}\, \|\omega-\omega'\|.
\end{eqnarray}
Here
$\tilde c(x)={\cal O}(1)$; its explicit form is unimportant and given in SM.
The property (\ref{eq:2}) allows us to use Pisier's theorem in CM \cite{Milman86} to show [SM (S4)] the following:\\
\\
\noindent {\it If $\omega=\{\omega_{(x,y)}\}$ is drawn randomly from an ensemble of disorder realizations with $\omega_{(x,y)}$ being independent Gaussian variables of zero mean and variance $\sigma^2$, then the probability for the deviation between $W(x;\omega)$ and its disorder mean ${\rm E}[W(x;\omega)]$ to exceed $\varepsilon$ satisfies}
\begin{equation}\label{eq:5}
    {\rm Pr}\left(|W(x;\omega)-{\rm E}[W(x;\omega)]|>\varepsilon\right)\leq 2e^{-\frac{2N\varepsilon^2}{(\tilde c(x)\pi\sigma)^2}}.
\end{equation}
Thus the concentration is strong for large $N$, i.e.,
\begin{eqnarray}
W(x;\omega)\approx {\rm E}[W(x;\omega)],\, for\,\, almost\,\, all\,\, \omega.
\label{eq:6}
\end{eqnarray}
It is important to remark that the factor $N$ in the sub-Gaussian bound of the concentration inequality (\ref{eq:5}) comes from the Lipschitz constant of $W(x;\omega)$, i.e., the coefficient on the right-hand side of the inequality (\ref{eq:2}).

\begin{figure}
\includegraphics[width=8.7cm] {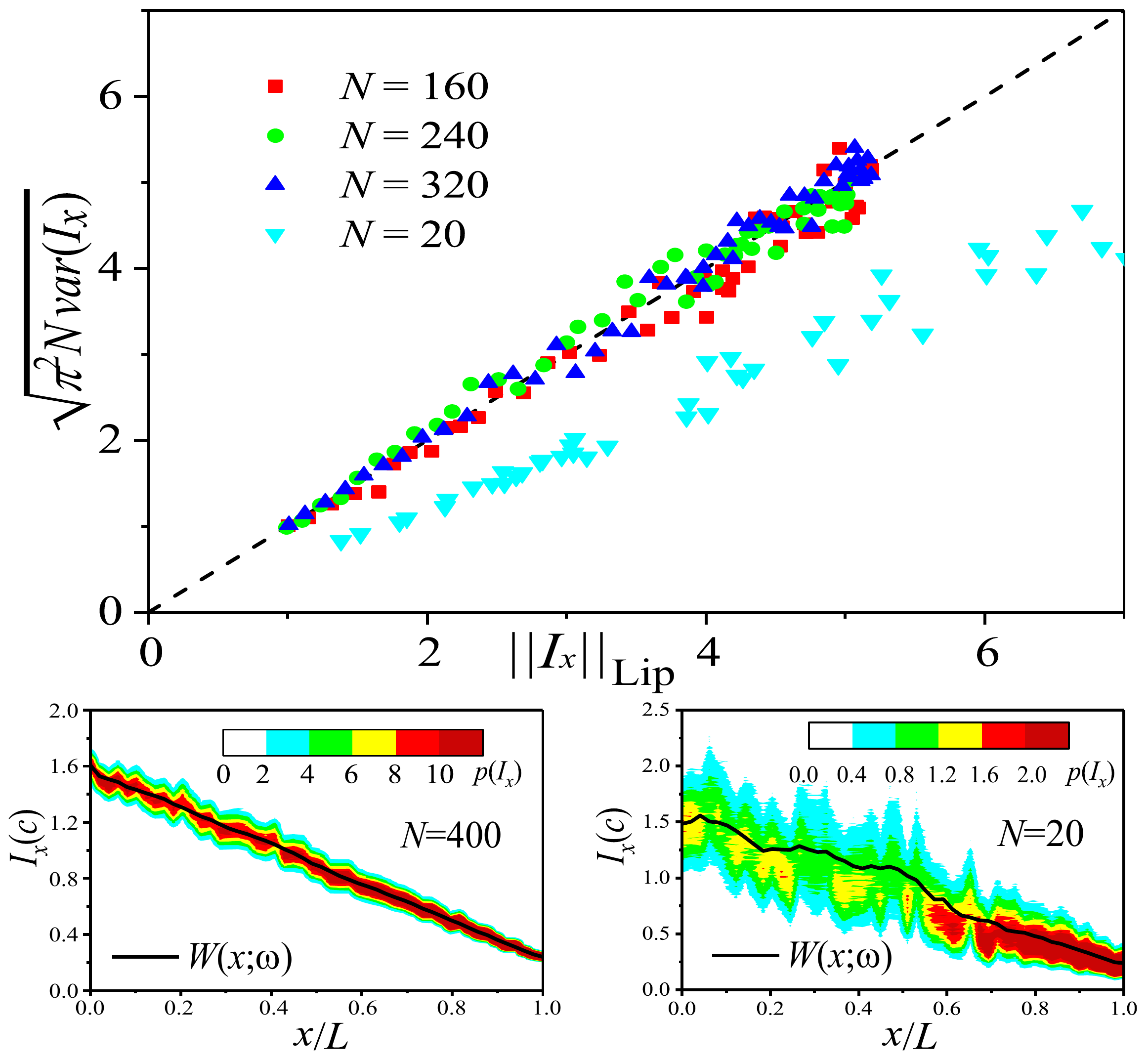}
\caption{Simulations show that in a single slab $\omega$ the profiles $I_x(c)$ for distinct $c$ concentrate around $W(x;\omega)$ (lower panels), and the data: $(\|I_x\|_{\rm Lip},\sqrt{\pi^2Nvar(I_x)})$ (symbols) collapse into a straight line of unit slope for distinct large $N$ while deviate from this line for small $N$ (upper panel). $L=50$}
\label{fig:1}
\end{figure}

To proceed to small $N$ we observe that the detailed structures of $\{W_{\tau_n}(x)\}$ enter into the inequality (\ref{eq:5}) only through the unimportant factor of $\tilde c(x)$. Thus we conjecture that the inequality applies in this case also. While its rigorous proof is beyond this work, we confirmed the conjecture numerically for $N$ being as small as $20$ [SM (S4)]. So Eq.~(\ref{eq:6}) holds for both large and small $N$.

Due to $I_x(c)\approx W(x;\omega)$ the disorder average of $I_x(c)$ ($c$ fixed) ${\rm E}[I_x(c)]\approx {\rm E}[W(x;\omega)]$. Together with Eq.~(\ref{eq:6}) this gives $W(x;\omega)\approx {\rm E}[I_x(c)]$.
As ${\rm E}[I_x(c)]$ is known to be the solution to the diffusion equation \cite{Sheng95,Rossum99,Akkermans}, $W(x;\omega)$ is a diffusive steady state (for almost all $\omega$), which decreases linearly in $x$. This result renders the transport mean free path $\ell$ well defined for single $\omega$ -- because for a diffusive steady state the total transmission $W(L;\omega)=\ell/L$ \cite{Rossum99} -- and identical to that defined for a disorder ensemble.

{\it Step 5 -- Wave-to-wave fluctuations in diffusive regime.} According to the criterion (\ref{eq:7}), weak wave-to-wave fluctuations ensure the emergence of a steady state universal with respect to $c$. To study these fluctuations we need to better understand $\|I_x\|_{\rm Lip}$. Let us start from large $N$. In SM (S1) we calculate Eq.~(\ref{eq:4}) and obtain for this case
\begin{eqnarray}\label{eq:3}
\begin{array}{c}
  \|I_x\|_{{\rm Lip}}=\int L_x(c;\omega) d\mu={\cal O}(1).
\end{array}
\end{eqnarray}
The profiles of $\|I_x\|_{{\rm Lip}}$ and $L_x$ are shown in Fig.~\ref{fig:4}. Surprisingly, we find that the profiles of $\|I_x\|_{{\rm Lip}}$ at distinct $N$ collapse into a single curve; moreover, the profile of $L_x$ is universal with respect to $c$ and $\omega$, and the universal curve is identical to that of $\|I_x\|_{{\rm Lip}}$.

This universality of $L_x$, together with Eq.~(\ref{eq:3}), implies the universality of $\|I_x\|_{{\rm Lip}}$ with respect to $\omega$. As shown in SM (S1), it even leads to an explicit expression of $\|I_x\|_{{\rm Lip}}^2$:
\begin{eqnarray}
\label{eq:14}
\begin{array}{c}
  \|I_x\|_{\rm Lip}^2=\frac{\pi^2}{N+1}\big(\sum_{n=1}^N\left(W_{\tau_n}(x)-W(x;\omega)\right)^2\\
  +\sum_{n\neq n'}^N|\sum_{a=1}^NE_{na}^*(x)E_{n'a}(x)|^2\big).
\end{array}
\end{eqnarray}
By using Eq.~(\ref{eq:14}) we find in SM (S5) that
\begin{equation}\label{eq:20}
    var(I_x)=\|I_x\|_{\rm Lip}^2/(\pi^2 N),
\end{equation}
and thus includes both incoherent and coherent contributions of eigenchannels, corresponding respectively to the first and second term in Eq.~(\ref{eq:14}). The incoherent (coherent) contribution dominates the back (front) part of the medium. At the output end, i.e., $x=L$, the second term in Eq.~(\ref{eq:14}) vanishes and the fluctuations include the incoherent contribution only.

For small $N$ the universality above is violated. But numerical calculations show that $\|I_x\|_{{\rm Lip}}={\cal O}(1)$ still holds (see the symbols corresponding to $N=20$ in the upper panel of Fig.~\ref{fig:1}). Due to this and $W(x;\omega)={\cal O}(1)$, for diffusive waves the criterion (\ref{eq:7}) can be readily satisfied.

{\it Numerical confirmations.} We put the theory into numerical tests. The methods of numerical experiments are described in SM (S7). First, we simulate wave propagation in a single slab for $10^4$ randomly chosen $c$. Simulations confirm that the profiles $I_x(c)$ concentrate around a linear decrease for both large and small $N$ (Fig.~\ref{fig:1}, lower panels). This also gives $\ell=13$ for single $\omega$ \cite{note3}. Moreover, from the distribution
$p(I_x)$ of wave-to-wave fluctuations we compute $var(I_x)$, and find that the relation (\ref{eq:20}) holds for large $N$ but is violated for small $N$ (Fig.~\ref{fig:1}, upper panel), as expected by our theory. Secondly, we simulate propagation of diffusive and localized waves in quasi $1$D media. We perform the statistics of wave-to-wave and sample-to-sample fluctuations of $I_x$: for the former we fix
the disorder realization $\omega$ and randomly choose $10^5$ incoming current amplitudes $c$ and for the latter do the opposite.
As shown in Fig.~\ref{fig:3}, for both diffusive and localized waves the distribution of wave-to-wave fluctuations displays a Gaussian tail, in agreement with the inequality (\ref{eq:1}). In contrast, the distribution of sample-to-sample fluctuations is much broader, which is exponential for diffusive waves and stretched-exponential for localized waves \cite{note1}.

Our theory provides new perspectives for the long-standing problem of the
emergence of irreversibility in individual systems. In particular, in SM (S6) we consider a generic hermitian (which is not essential) operator $\hat O$, and study its expectation value at the stationary scattering state $E$$($$x$,$y$$)$ (determined by the incoming wave $c$) is $O(c)$$\equiv$$\langle E|\hat O|E\rangle$$=$$\sum_{n,n'}$$c_n^*$$c_{n'}$$\langle E_n|\hat O|E_{n'}\rangle$. Recall that $E_n$ is the wave field $E_n$$($$x$,$y$$)$ of the $n$th eigenchannel. Repeating the analysis above we find
\begin{equation}\label{eq:13}
\begin{array}{c}
  O(c)\!\approx\!\frac{1}{N}\sum_{n=1}^N \langle E_n|\hat O|E_n\rangle\equiv {\bar O},\, for\, almost\, all\,c,
\end{array}
\end{equation}
if ${\bar O}$$\gg$$\frac{\|O\|_{{\rm Lip}}}{\sqrt{N}}$. This incoming
wave-independence of observables resembles
thermalization in closed systems \cite{Lebowitz06,Popescu06,Deutsch91,Srednicki94,Rigol08,Rigol16,Borhonovi16,von Neumann29}. But, as the systems here are open conceptual differences exist. Notably, bound states and equilibrium thermal ensembles
in closed systems are replaced respectively by stationary scattering states and $\frac{1}{N}$$\sum_{n}$$|E_n$$\rangle$$\langle$$E_n|$, which may be called the eigenchannel ensemble. In the future it is interesting to generalize Eq.~(\ref{eq:13}) to many-body systems, which would enable us to explore the relations between diffusive steady states and entanglement structures, but, unlike Ref.~\cite{Huse18}, requiring neither reservoirs nor disorder ensembles.

C.T. is grateful to J.-C. Garreau, F.-L. Lin, and Z. Q. Zhang, especially to A. Z. Genack and I. Guarneri for inspiring discussions and comments on the manuscript. This work is supported by the National Natural Science Foundation of China (No. 11535011 and No. 11747601).

\clearpage

\renewcommand{\thesection}{S\arabic{section}}
\renewcommand{\thesubsection}{\thesection.\arabic{subsection}}
\renewcommand{\theequation}{S\arabic{equation}}
\renewcommand{\thefigure}{S\arabic{figure}}

\setcounter{page}{1}
\setcounter{equation}{0}
\setcounter{figure}{0}

\begin{center}
{\bf Supplemental Materials}\\
by Ping Fang, Liyi Zhao, and Chushun Tian*
\end{center}

This appendix is written in a self-contained manner. The CM is introduced in great details. The technical details in quantitative (mathematical and numerical) analysis are presented, and extensive discussions on the results are made. The appendix is organized as follows:
\begin{itemize}
  \item Sec.~S0: It serves as an introduction of CM. We discuss some basics of CM, and introduce the mathematical results used in this work. Particular emphasis are put on their backgrounds and implications.
  \item Sec.~S1: We present a detailed analysis of the Lipschitz constant $\|I_x\|_{\rm Lip}$, which leads to the general result of
  Eq.~(\ref{eq:4}). Then we calculate this general result to obtain Eq.~(\ref{eq:3}). We further show Eq.~(\ref{eq:14}).
  \item Sec.~S2: We study the $2$D and $1$D eigenchannel structures in a slab. Comparisons with eigenchannel structures in quasi $1$D media are made.
  \item Sec.~S3: We study in details the Lipschitz continuity of $W(x;\omega)$ as a function of the disorder realization $\omega$, which leads to the inequality (\ref{eq:2}). Moreover, we derive analytically the expression of $\tilde c(x)$.
  \item Sec.~S4: We introduce a basic tool of CM, namely, Pisier's theorem. Armed with this theorem and the results obtained in Sec.~S3, we prove the concentration inequality (\ref{eq:5}). We further provide numerical evidences showing that it holds for both large and small $N$.
  \item Sec.~S5: We derive the relation (\ref{eq:20}). We perform further numerical studies of this relation.
  \item Sec.~S6: We discuss more general observables and derive Eq.~(\ref{eq:13}).
  \item Sec.~S7: We describe the method of simulations.
  \item Sec.~S8: We include the derivations of some integrals used in this appendix.
\end{itemize}

\noindent {\bf S0 Introduction of CM}\\

In this section we will illustrate that CM is a phenomenon of high-dimensional geometry, and introduce the background and implications of the mathematical results used in this work. In particular, we will show that the concentration of the uniform measure over the sphere $\mu$ lays a foundation for L$\acute{\rm e}$vy's lemma and the concentration of Gaussian measure over the Euclidean space $\gamma$ for Pisier's theorem. We will discuss that these two results are related to each other. By these discussions we hope that it will become clearer to the readers that CM opens ``completely new perspectives'' \cite{Boucheron13SM} of probability theory, and a CM-based probability theory differs from traditional probability theories such as the central limit theorem, with which physicists are familiar, in many fundamental aspects.\\
\\
\noindent {\bf S0.1 Concentration of uniform measure over the sphere}\\

\begin{figure}[h]
\includegraphics[width=8.7cm] {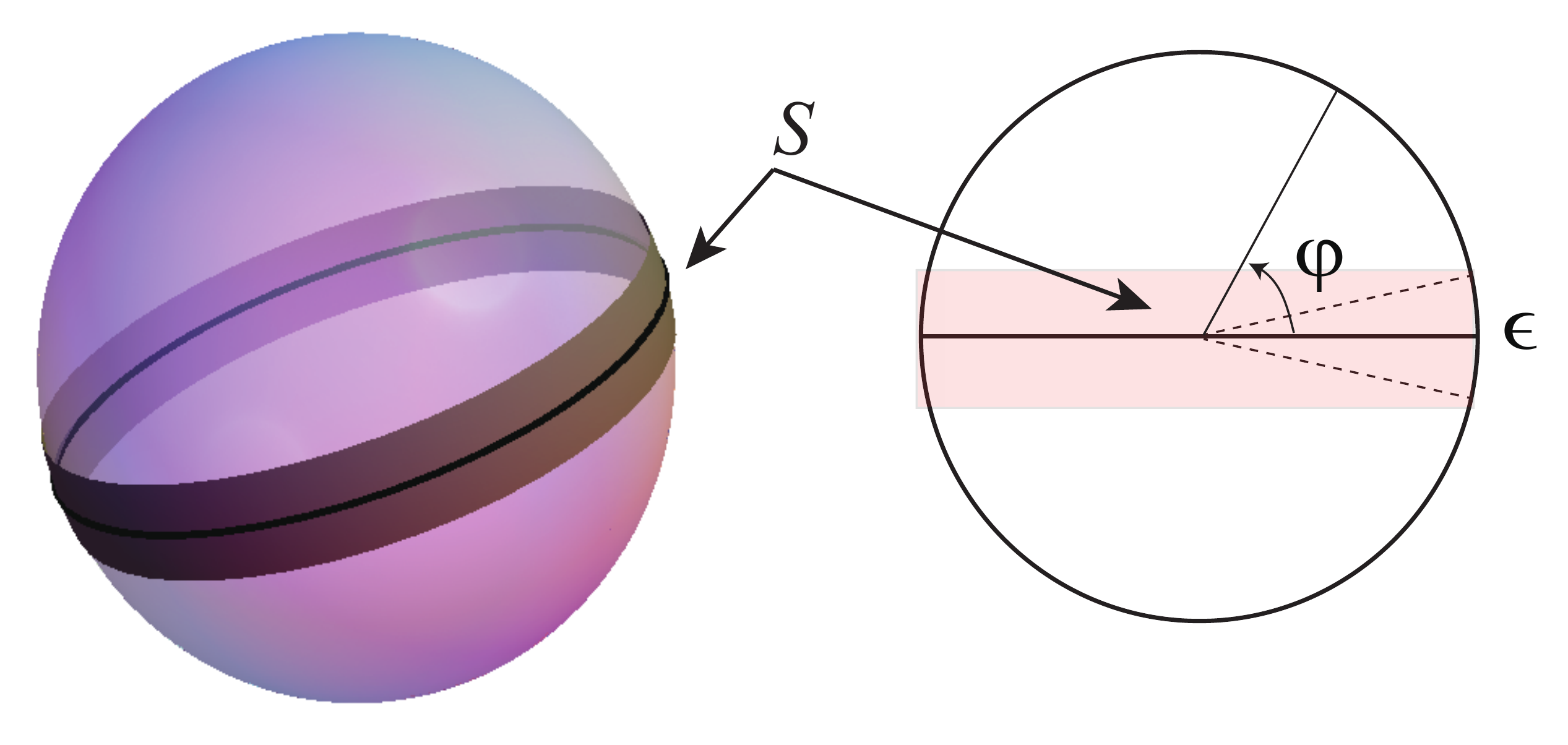}
\caption{A strip $S$ around an equator in the sphere (left) and its projection onto a $2$D plane (right). The latitude of the strip $\varphi \in [-\frac{\epsilon}{2}, \frac{\epsilon}{2}]$.}
\label{fig:S6}
\end{figure}

In this part we discuss in details a canonical example of CM. Let us take a unit sphere $S^{2N-1}$ in the Euclidean space $\mathds{R}^{2N}$ and normalize its area to unity. Consider a strip $S$ around and symmetric with respect to an equator (Fig.~\ref{fig:S6}). $S$ has a range of the latitude $\varphi$ between $-\frac{\epsilon}{2}$ and $\frac{\epsilon}{2}$. The north and south poles correspond the latitude of $-\frac{\pi}{2}$ and $\frac{\pi}{2}$, respectively. Using the parametrization (\ref{eq:S63})-(\ref{eq:S65}) introduced in Sec.~S8, we find that the area of the strip is
\begin{eqnarray}
\label{eq:S111}
  \mu(S)&=&\frac{\int_{-\frac{\epsilon}{2}}^{\frac{\epsilon}{2}} \sin^{2N-2}\varphi d\varphi}{\int_{-\frac{\pi}{2}}^{\frac{\pi}{2}} \sin^{2N-2}\varphi d\varphi}\nonumber\\
  &=&\frac{1}{\pi}\frac{(2N-2)!!}{(2N-3)!!}\int_{-\frac{\pi}{2}}^{\frac{\epsilon}{2}} \sin^{2N-2}\varphi d\varphi.
\end{eqnarray}

\noindent From this we find the value of $\epsilon$ corresponding to $\mu(S)=0.9$ so that $S$ covers $90\%$ of the entire area of the sphere. The result is shown in Table~\ref{tab:1}. From the table we see that as $N$ increases the entire area of the sphere concentrates more and more around the equator. Eventually for large $N$ the entire area almost concentrates on a very narrow strip around the equator.

\begin{table}[htbp]
\caption{$\epsilon$-values for $\mu(S)=0.9$}
\begin{tabular}{c|c|c|c|c|c}
\hline\hline
$N$ & $1$ & $2$& $20$ & $200$ & $2000$\\
\hline
$\epsilon/\pi$ &$0.90$ &$0.59$ &$0.17$& $0.05$&$0.02$\\
\hline\hline
\end{tabular}
\label{tab:1}
\end{table}

As most areas concentrate around the equator, a spherical cap slightly larger than a hemisphere, which has an area of exactly one half, covers almost the entire area of the sphere. This implies that as a spherical cap becomes bigger and bigger, the cap area undergoes a sharp transition once the cap boundary reaches an equator. This geometric phenomenon can be extended to a general set with an area $\geq \frac{1}{2}$, and has a more general and rigorous statement. Specifically, let $A$ be a subset of $S^{2N-1}$. Its $\varepsilon$-{\it enlargement}, denoted as $A_\varepsilon$, is defined as a set including all points which are at a geodesic distance $\leq\varepsilon$ from some point in $A$. (In words, any point in $A_\varepsilon$ is close to some point in $A$.) Then we have the following \cite{Milman86SM}:\\
\\
\noindent Theorem 0.1. {\it Let $A$ be a subset in $S^{2N-1}$, whose area $\mu(A)\geq\frac{1}{2}$, and $N$ be an arbitrary natural number. Then}
\begin{equation}
    \mu(A_\varepsilon)\geq 1-e^{-(N-1)\varepsilon^2}.
    \label{eq:S113}
\end{equation}
\noindent {\it Remarks.} (i) According to this theorem, the entire area of the sphere concentrates around a strip around the equator with a width $\sim \frac{1}{\sqrt{N}}$ for large $N$. (ii) The theorem is the formalization of the concentration of the uniform measure $\mu$ over the sphere. (iii) To prove the theorem one needs to use the {\it isoperimetric inequality} discovered by L$\acute{\rm e}$vy \cite{Milman86SM}. We will not give its rigorous statement here. Instead, we explain the inequality in words. That is, in a sphere the circle enclosing a spherical cap is the curve of given length enclosing the largest area.

Next, we would like to probe the concentration of the measure $\mu$. Consider a real-valued function $f$$:$$S^{2N-1}$$\rightarrow$$\mathds{R}$ which has the Lipschitz continuity, with the Lipschitz constant
\begin{eqnarray}
\|f\|_{{\rm Lip}}\equiv{\rm sup}_{c,c'}\frac{|f(c)-f(c')|}{\|c-c'\|}<\infty.
\label{eq:S114}
\end{eqnarray}
Recall that $\|\cdot\|$ is the Euclidean metric in $\mathds{R}^{2N}$. Furthermore, we define the {\it median} or {\it L$\acute{e}$vy's mean} of $f$, denoted as $M[f]$, as the number such that both $\mu(f(c)\leq M[f])\geq \frac{1}{2}$ and $\mu(f(c)\geq M[f])\geq \frac{1}{2}$ are satisfied. Then by using Theorem 0.1 one can readily prove \cite{Milman86SM} the following:\\
\\
\noindent Corollary 0.2. {\it Let $f:S^{2N-1}\rightarrow\mathds{R}$ be Lipschitz. Then for any $\varepsilon>0$,
\begin{equation}
    {\rm Pr}(|f(c)-M[f]|\geq \varepsilon)\leq 2e^{-\frac{(N-1)\varepsilon^2}{\|f\|_{\rm Lip}^2}}.
    \label{eq:S115}
\end{equation}
}

\noindent {\it Remarks.} (i) This corollary is similar to L$\acute{\rm e}$vy's lemma introduced in the paper. It implies the concentration of function $f$ around the median of $M[f]$, and thus is another version of L$\acute{\rm e}$vy's lemma. We see again that Lipschitz functions serve as nice observables to probe the concentration of measure $\mu$ through the concentration of functions around the constant value $M[f]$. (ii) An important concept follows from the corollary. That is, the sub-Gaussian tail of the distribution of $f$ inherits from Theorem 0.1 or more precisely the isoperimetric inequality, and thus the sub-Gaussian nature of the distribution is of purely geometric origin and not related to the central limit theorem. (iii) $f(c)$ can have a very complicated dependence on $c$. (iv) However, $M[f]$ differs from the mean: $\int fd\mu$ in general. Due to this the corollary is not useful to us in practice.

A natural question is: how can we obtain L$\acute{\rm e}$vy's lemma used in the paper, which gives the concentration of $f$ around the mean $\int fd\mu$? To answer this question we make the following observation. We parametrize $N$ complex coefficients $c_n$ ($n=1,2,\cdots, N$), which satisfy $\sum_{n=1}^N |c_n|^2=1$ and thus constitute the coordinate of $S^{2N-1}$, as:
\begin{eqnarray}\label{eq:S58}
    c_n=\frac{a'_n+ib'_n}{\sqrt{\sum_{n=1}^N((a'_n)^2+(b'_n)^2)}},\quad a'_n, b'_n\in \mathds{R}.
\end{eqnarray}
Then, it is well known \cite{Milman86SM} that if the $2N$ real variables $a'_n, b'_n$ are independent standard normal random variables, then $c\equiv(c_1,c_2,\cdots,c_N)$ is distributed uniformly over $S^{2N-1}$. This motivates us to investigate the concentration of Gaussian measures $\gamma$ over the Euclidean space of general dimensions. This is the purpose of the next part. In doing so we will sketch how to obtain L$\acute{\rm e}$vy's lemma used in the paper.\\
\\
\noindent {\bf S0.2 Concentration of Gaussian measure over the Euclidean space}\\

The discussions above are built upon very few notions, namely, the geometric body $S^{2N-1}$ equipped with the geodesic metric and the uniform measure $\mu$. It is possible to extend the discussions to general geometric body equipped with appropriate metric and measure. We refer the general discussions to standard mathematical literatures \cite{Milman86SM,Ledoux01SM}. Here we study in details the Euclidean space $\mathds{R}^k$ of general dimension $k$, equipped with the Euclidean metric $\|\cdot\|$ and
the standard Gaussian probability measure:
\begin{equation}\label{eq:S116}
    d\gamma\equiv(2\pi)^{-\frac{k}{2}}e^{-\frac{\|\varpi\|^2}{2}}d\varpi,\quad \varpi\in\mathds{R}^k,
\end{equation}
which we used in this work.

For this measure, there exists a concentration phenomenon similar to what occurs to the uniform measure over the sphere. That is, a set slightly larger than the half space, which has a Gaussian measure of exactly one half, covers most part of the entire measure of the space, despite that the latter is noncompact. With the introduction of $\varepsilon$-enlargement $A_\varepsilon$ (in the same way as before, except that the geodesic distance is replaced by the Euclidean one), this phenomenon can be formalized as the following \cite{Ledoux01SM}:\\
\\
\noindent Theorem 0.3. {\it Let $A$ be a subset in $\mathds{R}^k$, whose Gaussian measure $\gamma(A)\geq\frac{1}{2}$, and $k$ be an arbitrary natural number. Then}
\begin{equation}
    \gamma(A_\varepsilon)\geq 1-e^{-\varepsilon^2/2}.
    \label{eq:S117}
\end{equation}
\noindent{\it Remark.} Compared with Theorem 0.1 we find that this CM phenomenon has a striking feature: it is {\it dimension free}, i.e., the right-hand side of the inequality (\ref{eq:S117}) has no $k$-dependence.

As before, the concentration of the measure $\gamma$ can be probed by nice observables. To be specific, we take any real-valued function $f$$:$$\mathds{R}^k$$\rightarrow$$\mathds{R}$ which has the Lipschitz continuity, with the Lipschitz constant
\begin{eqnarray}
\|f\|_{{\rm Lip}}\equiv{\rm sup}_{\varpi,\varpi'}\frac{|f(\varpi)-f(\varpi')|}{\|\varpi-\varpi'\|}<\infty.
\label{eq:S118}
\end{eqnarray}
Then we have the following \cite{Milman86SM,Ledoux01SM}:\\
\\
\noindent (Pisier) Theorem 0.4. {\it Let $f: \mathds{R}^k\rightarrow \mathds{R}$ be Lipschitz and $\mathds{R}^k$ be equipped with the standard Gaussian measure $\gamma$. Then for any $\varepsilon>0$,
\begin{equation}\label{eq:S119}
    {\rm Pr}\left(\left|f(\varpi)-{\rm E}[f]\right|>\varepsilon\right)\leq 2e^{-\frac{2\varepsilon^2}{(\|f\|_{\rm Lip}\pi)^2}},
\end{equation}
where the expectation value ${\rm E}[f]=\int fd\gamma$.}\\

\noindent{\it Remarks.} (i) Unlike Corollary 0.2, this theorem refers to the concentration around the mean, rather than the median. (ii) Inheriting from Theorem 0.3 the concentration of $f$ is dimension free also, i.e., the right-hand side of the concentration inequality (\ref{eq:S119}) has no $k$-dependence. This also shows that the sub-Gaussian nature of the distribution of $f(\varpi)$ is purely geometric and not related to the central limit theorem. (iii) $f(\varpi)$ can have a very complicated dependence on $\varpi$.

Now consider $k$$=$$2$$N$ and $\varpi$$\equiv$$($$a'_1$,$b'_1$,$\cdots$,$a'_N$,$b'_N)$, where $a'_n$,$b'_n$ are independent Gaussian variables parametrizing $c_n$ in the way shown by Eq.~(\ref{eq:S58}). Then, by using Theorem 0.4 one can prove L$\acute{\rm e}$vy's lemma used in the paper \cite{Milman86SM}:\\
\\
\noindent (L$\acute{\rm e}$vy) Lemma 0.5. {\it Let $f: S^{2N-1}\rightarrow \mathds{R}$ be Lipschitz. Then for any $\varepsilon>0$,
\begin{equation}\label{eq:S120}
    {\rm Pr}\left(\left|f(c)-\int f d\mu\right|>\varepsilon\right)\leq 2e^{-
\frac{\delta\varepsilon^2 N}{\|f\|_{{\rm Lip}}^2}},
\end{equation}
where $\delta$ is some positive absolute constant.}\\

The studies of wave-to-wave fluctuations are based on L$\acute{\rm e}$vy's lemma. As shown above, this lemma is a nontrivial application of Pisier's theorem. Later on, we will see that Pisier's theorem is very useful also in the studies of sample-to-sample fluctuations of $W(x;\omega)$ and $L_x(c;\omega)$.\\
\\
\noindent {\bf S1 Calculations and properties of $\boldsymbol{\|I_x\|_{\rm Lip}}$}\\

This section is divided into five subsections. In Sec.~S1.1 we show that if $I_x: S^{2N-1}\rightarrow \mathds{R}$ is continuously differentiable, i.e.,
\begin{equation}\label{eq:S76}
    {\rm sup}_c\|\nabla_c I_x\|<\infty,
\end{equation}
where $\nabla_c$ is the orthographic projection of $\nabla$ in $\mathds{R}^{2N}$ onto $S^{2N-1}$, then $I_x(c)$ is Lipschitz. In other words, the continuous differentiability is stronger than the Lipschitz continuity. In doing so, we connect the Lipschitz constant with respect to distinct metrics, namely, the geodesic metric and the Euclidean metric (in $\mathds{R}^{2N}$). In Sec.~S1.2 by deriving Eq.~(\ref{eq:4}) we confirm that the continuous differentiability, namely, the condition (\ref{eq:S76}) is indeed satisfied. In Sec.~S1.3 we perform numerical analysis and establish the universality of $\|I_x\|_{\rm Lip}$ (with respect to the Euclidean metric) at large $N$. In Sec.~S1.4 we explain this universality by CM. In Sec.~S1.5 we use this universality to derive Eq.~(\ref{eq:14}).\\
\\
\noindent {\bf S1.1 Results of continuous differentiability}\\

We start from showing that the continuous differentiability implies the Lipschitz continuity:\\

\noindent{Lemma 1.1.} {\it If $I_x(c)$ is continuously differentiable, i.e., the condition (\ref{eq:S76}) holds, then}
\begin{equation}
    |I_x(c)-I_x(c')|\leq \frac{\pi}{2}{\rm sup}_c\|\nabla_c I_x\|\|c-c'\|.
\label{eq:S112}
\end{equation}

{\it Proof.} Suppose that $\gamma (c,c')$ is the shorter geodesic connecting $c,c'\in S^{2N-1}$.  Then
\begin{equation}\label{eq:17}
    |I_x(c)-I_x(c')|=\left|\int_{\gamma (c,c')}ds \frac{\partial I_x}{\partial s}\right|\leq \int_{\gamma (c,c')}ds \left|\frac{\partial I_x}{\partial s}\right|,
\end{equation}
where $ds$ is the differential element of $\gamma (c,c')$. Because of $|\partial I_x/\partial s|\leq \|\nabla_c I_x\|$,
we have
\begin{eqnarray}\label{eq:19}
    \int_{\gamma (c,c')}ds \left|\frac{\partial I_x}{\partial s}\right|&\leq& \int_{\gamma (c,c')}ds \|\nabla_c I_x\|\nonumber\\
    &\leq& {\rm sup}_c\|\nabla_c I_x\| \int_{\gamma (c,c')}ds.
\end{eqnarray}
Note that $\int_{\gamma (c,c')}ds$ is the geodesic distance of $\gamma$. It satisfies
\begin{equation}\label{eq:27}
\int_{\gamma (c,c')}ds\leq \frac{\pi}{2}\|c-c'\|.
\end{equation}
Combining the inequalities (\ref{eq:17}), (\ref{eq:19}) and (\ref{eq:27}), we prove the lemma. $\Box$

{\it Remarks.} (i) That the continuous differentiability implies the Lipschitz continuity is actually very general. In particular, one may replace $S^{2N-1}$ by arbitrary space $\mathscr{C}$ (equipped with certain metric). (ii) The lemma shows that
\begin{equation}\label{eq:S77}
    \|I_x\|_{{\rm Lip}}\leq \frac{\pi}{2}{\rm sup}_c\|\nabla_c I_x\|.
\end{equation}
On the other hand, because of
\begin{eqnarray}
    \lim_{c'\rightarrow c}\frac{|I_x(c)-I_x(c')|}{\|c-c'\|}\leq \|\nabla_c I_x\|\leq {\rm sup}_{c}\|\nabla_c I_x\|,
    \label{eq:S78}
\end{eqnarray}
we have
\begin{eqnarray}
    \|I_x\|_{{\rm Lip}}\geq {\rm sup}_{c}\|\nabla_c I_x\|.
    \label{eq:S101}
\end{eqnarray}
The inequalities (\ref{eq:S77}) and (\ref{eq:S101}) imply that
\begin{equation}\label{eq:S80}
    \|I_x\|_{{\rm Lip}}=a{\rm sup}_c\|\nabla_c I_x\|,\quad 1\leq a\leq \pi/2,
\end{equation}
where $a$ is some numerical constant. Because $a$ does not grow with $N$ and is close to unity, it does not affect any results obtained from L$\acute{\rm e}$vy's lemma, except slightly modifying the absolute constant $\delta$ in the sub-Gaussian tail. Thus we do not pay attention to its exact value any more and set $a=\pi/2$, i.e.,
\begin{equation}\label{eq:S51}
    \|I_x\|_{{\rm Lip}}=\frac{\pi}{2}{\rm sup}_c\|\nabla_c I_x\|\equiv {\rm sup}_c L_x(c;\omega).
\end{equation}
This provides an analytic formula of $\|I_x\|_{{\rm Lip}}$, which we will use in subsequent mathematical and numerical analysis.\\
\\
\noindent {\bf S1.2 Derivations of Eq.~(\ref{eq:4}): confirming continuous differentiability}\\

In this part we use Eq.~(\ref{eq:S51}) to derive Eq.~(\ref{eq:4}). Let $c_n=a_n+ib_n$, where $a_n,b_n\in \mathds{R}$. By using the definition (\ref{eq:10}) of $I_x(c)$, we obtain
\begin{eqnarray}\label{eq:S54}
    \|\nabla I_x\|^2&=&\sum_{n=1}^N \left(\left(\frac{\partial I_x}{\partial a_n}\right)^2+\left(\frac{\partial I_x}{\partial b_n}\right)^2\right)\nonumber\\
    &=&4\sum_{n=1}^N \frac{\partial I_x}{\partial c_n}\frac{\partial I_x}{\partial c_n^*},
\end{eqnarray}
where
\begin{eqnarray}\label{eq:30}
    \frac{\partial I_x}{\partial c_n}&=&\sum_{n'=1}^N \!c_{n'}^* E_{n'}^\dagger(x)\cdot E_{n}(x),\nonumber\\
    \frac{\partial I_x}{\partial c_n^*}&=&\sum_{n'=1}^N \!c_{n'} E_{n}^\dagger(x)\cdot E_{n'}(x).
\end{eqnarray}
Denote $\boldsymbol{n}\equiv (a_1,b_1,\cdots,a_N,b_N)$. It is easy to see that $\boldsymbol{n}$ is the radial direction of $S^{2N-1}$. Moreover, because of $\sum_{n=1}^N(a_n^2+b_n^2)=1$ it is a unit vector. As a result,
\begin{eqnarray}\label{eq:S55}
    \boldsymbol{n}\cdot\nabla I_x&=&\sum_{n=1}^N \left(a_n\frac{\partial I_x}{\partial a_n}+b_n\frac{\partial I_x}{\partial b_n}\right)\nonumber\\
    &=&\sum_{n=1}^N \left(c_n\frac{\partial I_x}{\partial c_n}+c_n^*\frac{\partial I_x}{\partial c_n^*}\right).
\end{eqnarray}
By the definition (\ref{eq:10}), we find
\begin{eqnarray}\label{eq:S56}
    \sum_{n=1}^Nc_n\frac{\partial I_x}{\partial c_n}=\sum_{n=1}^Nc_n^*\frac{\partial I_x}{\partial c_n^*}=I_x.
\end{eqnarray}
Substituting it into Eq.~(\ref{eq:S55}) gives
\begin{eqnarray}\label{eq:S57}
    \boldsymbol{n}\cdot\nabla I_x=2I_x.
\end{eqnarray}
Combining Eqs.~(\ref{eq:S54}) and (\ref{eq:S57}) we find
\begin{eqnarray}\label{eq:29}
    L_x(c;\omega)&=&\frac{\pi}{2}\|\nabla_c I_x\|\nonumber\\
    &=&\frac{\pi}{2}\sqrt{\|\nabla I_x\|^2-|\boldsymbol{n}\cdot\nabla I_x|^2}\nonumber\\
    &=&\pi\sqrt{\sum_{n=1}^N\frac{\partial I_x}{\partial c_n}\frac{\partial I_x}{\partial c_n^*}-I_x^2}.
\end{eqnarray}
Equations (\ref{eq:S51}), (\ref{eq:30}) and (\ref{eq:29}) give Eq.~(\ref{eq:4}). According to Eq.~(\ref{eq:4}), the condition (\ref{eq:S76}) is satisfied, i.e., $I_x$ is continuously differentiable.\\
\\
\noindent {\bf S1.3 Results of Eq.~(\ref{eq:4})}\\

It is difficult to calculate Eq.~(\ref{eq:4}) explicitly and analytically, especially in view of that $\omega$ is fixed. However, to calculate Eq.~(\ref{eq:4}) numerically is relatively easy. The point is that, owing to the analytic expression of $L_x(c;\omega)$, we do not need to perform numerical differentiation which is technically demanding. To be specific, given $\omega$ we can simulate Eq.~(\ref{eq:15}) by using the method described in Sec.~S7 and obtain the set of $\{E_n(x)\}$. Upon substituting the result into Eq.~(\ref{eq:4}) we obtain $L_x(c;\omega)$ and $\|I_x\|_{\rm Lip}$. In this part we put this numerical scheme into practice and calculate Eq.~(\ref{eq:4}) explicitly. Surprisingly, we will show that for large $N$, $\|I_x\|_{\rm Lip}$ displays universalities, and even stronger results hold for $L_x(c;\omega)$.

Given $N$ we draw $c$ randomly from a uniform distribution over $S^{2N-1}$ by using the algorithm described in Sec.~S7. Then we simulate the wave propagation in a fixed disorder realization $\omega$ (the details are described in Sec.~S7.), and calculate $\|\nabla_c I_x\|$ or equivalently $L_x(c;\omega)$ by using the method described above.
This gives a depth profile of $\|\nabla_c I_x\|$. We repeat the numerical experiment for $10^4$ randomly chosen $c$, and thus have $
10^4$ profiles in total. We have checked that the number of $
10^4$ is large enough to ensure that numerical results converge. Finally, the procedures are repeated for several distinct $N$.

\begin{figure}
\includegraphics[width=8.7cm] {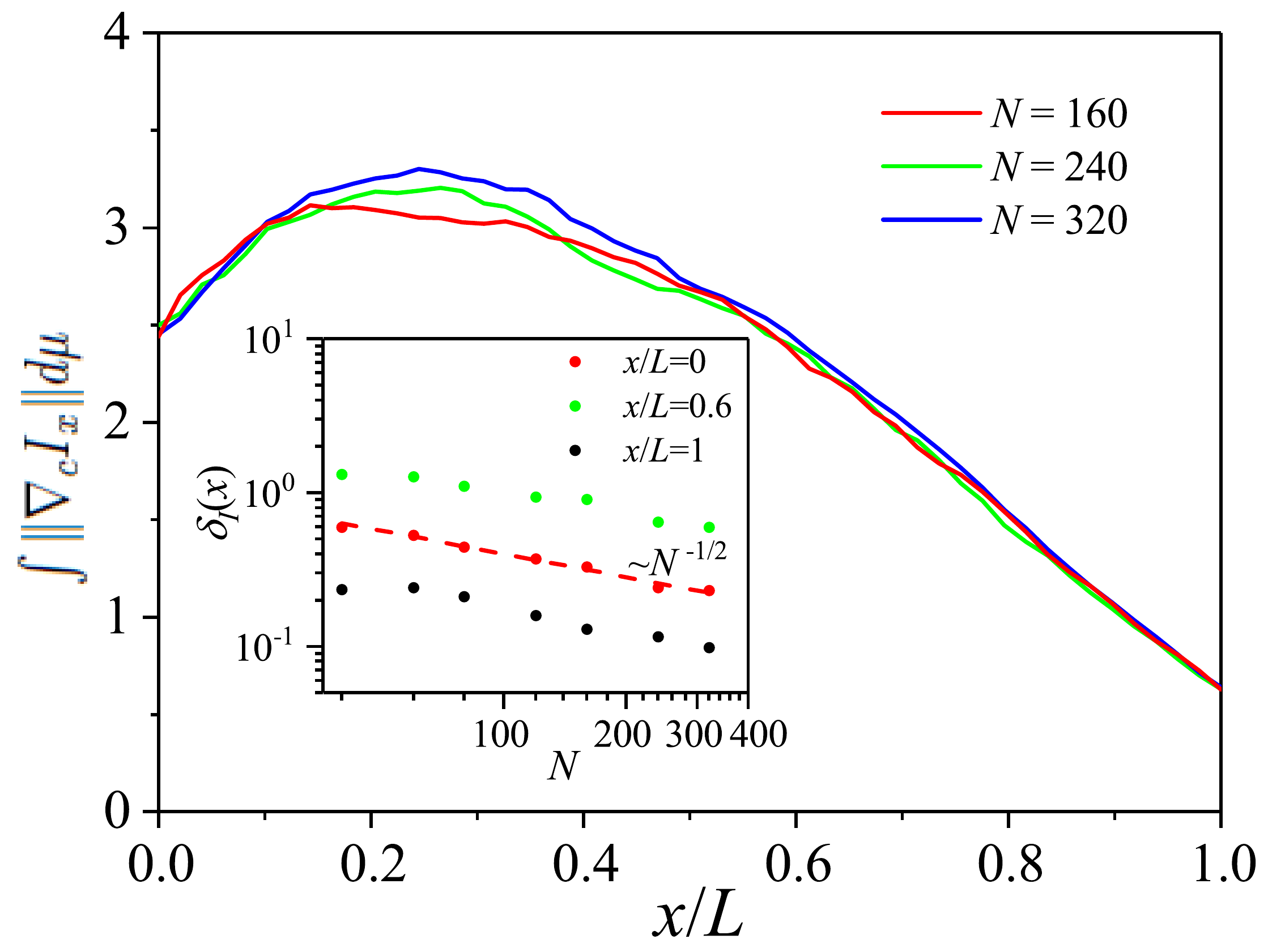}
\caption{Main panel: Simulations show that the profiles of $\int \|\nabla_c I_x\|d\mu$ corresponding to $N=160,240,320$ collapse into a single curve, and $\int \|\nabla_c I_x\|d\mu={\cal O}(1)$ at every $x$. Inset: Simulations show that given $x$ the maximal deviation $\delta_I(x)$ from $\int \|\nabla_c I_x\|d\mu$ decreases with $N$ as $N^{-1/2}$.}
\label{fig:S1}
\end{figure}

To obtain the depth profile of $\|I_x\|_{\rm Lip}$, we perform the numerical analysis of the profiles of $\|\nabla_c I_x\|$ in the following way. (i) We average the profiles corresponding to the same $N$. The average profile, $\int \|\nabla_c I_x\|d\mu
$, is shown in the main panel of Fig.~\ref{fig:S1} for three different $N$. We see that these three profiles collapse into a single curve which is continuous in $x$, and the value of $\int \|\nabla_c I_x\|d\mu$ is order of unity at every $x$. (ii) We calculate the maximal deviation $\delta_I(x)$ from the average value $\int \|\nabla_c I_x\|d\mu$ for distinct $x$. As shown in the inset of Fig.~\ref{fig:S1}, $\delta _I(x)$ scales with $N$ as
\begin{equation}\label{eq:S49}
    \delta _I(x)\sim \frac{1}{\sqrt{N}},\quad \forall\, x\in [0,L].
\end{equation}
This implies that for large $N$, the two profiles, $\|\nabla_c I_x\|$ and $\int \|\nabla_c I_x\|d\mu$, converge, giving
\begin{equation}\label{eq:S50}
    \|\nabla_c I_x\|=\int \|\nabla_c I_x\|d\mu,
\end{equation}
i.e., $\|\nabla_c I_x\|$ is independent of $c$.

By using Eq.~(\ref{eq:4}) namely Eq.~(\ref{eq:S51}) and Eq.~(\ref{eq:S50}) we find
\begin{equation}\label{eq:S1}
    \|I_x\|_{{\rm Lip}}=\frac{\pi}{2} \int \|\nabla_c I_x\|d\mu.
\end{equation}
Since as shown in (i) the right-hand side has no $N$-dependence and is order of unity, Eq.~(\ref{eq:3}) follows. The curves in Fig.~\ref{fig:S1}, when multiplied by $\pi/2$, give the three $\|I_x\|_{\rm Lip}$-curves in Fig.~\ref{fig:4}.

The finding of Eq.~(\ref{eq:S50}) motivates us to perform a careful study of the universalities of $\|\nabla_c I_x\|=\frac{2}{\pi}L_x(c;\omega)$. We perform simulations for large $N(=800)$. Surprisingly, we find that $L_x(c;\omega)$ is independent not only of $c$, but also of $\omega$, as shown in Fig.~\ref{fig:4}. In other words, $L_x(c;\omega)$ gives a universal curve parametrized by $x$ for very large $N$.\\
\\

\noindent {\bf S1.4 Universality of $\boldsymbol{L_x(c;\omega)}$ from CM}\\

In this part we employ CM to explain the universality of $L_x(c;\omega)$ with respect to $c$ and $\omega$. We stress that this is not a rigorous mathematical proof, which is far beyond the present work.

First of all, we fix $\omega$. From Eq.~(\ref{eq:4}) or Eq.~(\ref{eq:29}) we see that $L_x(c;\omega)$ is continuously differentiable, i.e.,
\begin{equation}\label{eq:S106}
    {\rm sup}_c\|\nabla_c L_x\|<\infty,
\end{equation}
and thus is Lipschitz. Let the Lipschitz constant of $L_x(c;\omega)$ ($\omega$ fixed) be $\|L_x(\omega)\|_{\rm Lip}$. By making use of Lemma 0.5 we obtain
\begin{eqnarray}\label{eq:S107}
  {\rm Pr}\left(\left|L_x(c;\omega)-\int L_x(c;\omega)d\mu\right|>\varepsilon\right)\leq 2e^{-
    \frac{\delta\varepsilon^2 N}{\|L_x(\omega)\|_{{\rm Lip}}^2}}.\nonumber\\
\end{eqnarray}
From this we see that provided
\begin{equation}\label{eq:S108}
    \int L_x(c;\omega)d\mu\gg \|L_x(\omega)\|_{{\rm Lip}}/\sqrt{N},
\end{equation}
strong concentration around the mean: $\int L_xd\mu$ results. This gives the universality with respect to $c$.

Next, we fix $c$. From Eq.~(\ref{eq:4}) or Eq.~(\ref{eq:29}) we may expect that $L_x(c;\omega)$ is continuously differentiable with respect to $\omega\equiv\{\omega_{(x,y)}\}$, when the medium space is discretized into $M$ lattice points. Thus we have
\begin{eqnarray}\label{eq:S109}
  |L_x(c;\omega)-L_x(c;\omega')|\leq
  \|L_x(c)\|_{{\rm Lip}}\, \|\omega-\omega'\|,
\end{eqnarray}
where $\|L_x(c)\|_{{\rm Lip}}$ is the corresponding Lipschitz constant. The ensemble of disorder realizations $\omega$, by definition, is described by a Gaussian probability measure on $\mathds{R}^M$ of zero mean and variance $\sigma^2$. Let $\varpi$ in (Pisier's) Theorem 0.4. be $\omega$, $f(\varpi)$ be $L_x(c;\omega)$, and $k=M$. After rescaling we obtain from this theorem
\begin{equation}\label{eq:S110}
    {\rm Pr}\left(|L_x(c;\omega)-{\rm E}[L_x(c;\omega)]|>\varepsilon\right)\leq 2e^{-\frac{2\varepsilon^2}{(\pi\|L_x(c)\|_{{\rm Lip}}\sigma)^2}}.
\end{equation}
It is important to note that for this concentration inequality the exponent of the sub-Gaussian tail bound has no explicit $N$-dependence. Now, if we assume that $\|L_x(c)\|_{{\rm Lip}}$ decays with $N$, then for sufficiently large $N$ we have strong concentration of $L_x(c;\omega)$ around the disorder average: ${\rm E}[L_x(c;\omega)]$. This explains the universality of $L_x(c;\omega)$ with respect to $\omega$.

Summarizing, the non-rigorous discussions above indicate that the universalities of $L_x(c;\omega)$ may have deep connections to CM.\\
\\
\noindent {\bf S1.5 Derivations of Eq.~(\ref{eq:14})}\\

In this part we study the consequences of the universalities of $L_x(c;\omega)$. By Jensen's inequality we have:
\begin{eqnarray}
\label{eq:S105}
  \|I_x\|_{{\rm Lip}}^2=\left(\int L_x(c;\omega)d\mu\right)^2\leq \int L_x^2(c;\omega)d\mu
\end{eqnarray}
in general. With the universality of $L_x(c;\omega)$ with respect to $c$ taken into account, a stronger result follows, i.e.,
\begin{equation}\label{eq:S88}
    \|I_x\|_{{\rm Lip}}^2=\int L_x^2(c;\omega)d\mu.
\end{equation}
Now we show that this gives Eq.~(\ref{eq:14}).

Substituting Eq.~(\ref{eq:4}) into it we obtain
\begin{eqnarray}
\label{eq:S89}
\frac{\|I_x\|_{{\rm Lip}}^2}{\pi^2} = \sum_{n,n',m'}^N\!\! \left(\int c_{m'}^*c_{n'} d\mu\right) E_{m'}^\dagger\cdot E_{n}E_{n}^\dagger\cdot E_{n'}\quad\nonumber\\
  -\!\sum_{n,n',m,m'}^N\!\! \left(\int c_{n}^*c_{n'}c_{m}^*c_{m'} d\mu\right) E_{n}^\dagger\cdot E_{n'}E_{m}^\dagger\cdot E_{m'}.\,\,
\end{eqnarray}
According to the integral formula (\ref{eq:S79}) derived in Sec.~S8, for the integral $\int c_{m'}^*c_{n'} d\mu$ not to vanish it is necessary $m'=n'$. For the integral $\int c_{n}^*c_{n'}c_{m}^*c_{m'} d\mu$ not to vanish, it is necessary that one of the following conditions: (i) $n=n', m=m' (n\neq m)$; (ii) $n=m', m=n' (n\neq m)$; (iii) $n=n'=m=m'$ must be satisfied. As a result,
\begin{eqnarray}\label{eq:S90}
    &&\int c_{n}^*c_{n'}c_{m}^*c_{m'} d\mu\nonumber\\
    &=&\int |c_{n}|^2|c_{m}|^2 d\mu(\delta_{nn'}\delta_{mm'}+\delta_{nm'}\delta_{mn'})(1-\delta_{nm})\nonumber\\
    &+&\int |c_{n}|^4 d\mu \delta_{nn'}\delta_{mm'}\delta_{nm}.
\end{eqnarray}
Using the integral formulae (\ref{eq:S60}) and (\ref{eq:S61}) derived in Sec.~S8, we obtain
\begin{eqnarray}\label{eq:S91}
    &&\int c_{n}^*c_{n'}c_{m}^*c_{m'} d\mu\nonumber\\
    &=&\frac{1}{N(N+1)}(\delta_{nn'}\delta_{mm'}+\delta_{nm'}\delta_{mn'})(1-\delta_{nm})\nonumber\\
    &+&\frac{2}{N(N+1)}\delta_{nn'}\delta_{mm'}\delta_{nm}.
\end{eqnarray}
Using this result we reduce Eq.~(\ref{eq:S89}) to
\begin{eqnarray}
\label{eq:S92}
\frac{\|I_x\|_{{\rm Lip}}^2}{\pi^2} = \frac{1}{N}\sum_{n,m=1}^N |E_{n}^\dagger\cdot E_{m}|^2\quad\quad\quad\quad\quad\quad\quad\quad\quad\quad\nonumber\\
  -\frac{1}{N(N+1)}\sum_{n\neq m}^N \left(E_{n}^\dagger\cdot E_{n}E_{m}^\dagger\cdot E_{m}+|E_{n}^\dagger\cdot E_{m}|^2\right)\nonumber\\
  -\frac{2}{N(N+1)}\sum_{n=1}^N |E_{n}^\dagger\cdot E_{n}|^2.\quad\quad\quad\quad\quad\quad\quad\quad\quad\,\,\,\,
\end{eqnarray}
From this we obtain
\begin{eqnarray}
\label{eq:S121}
  \|I_x\|_{\rm Lip}^2&=&\frac{\pi^2}{N+1}\bigg(\sum_{n=1}^N\left(W_{\tau_n}(x)-W(x;\omega)\right)^2\nonumber\\
  &&+\sum_{n\neq n'}^N|E_{n}^\dagger(x)\cdot E_{n'}(x)|^2\bigg).
\end{eqnarray}
after simple algebra. Recall that $E_n(x)\equiv \{E_{na}(x)\}$ and $a$ labels the ideal waveguide modes. We find that Eq.~(\ref{eq:S121}) is equivalent to Eq.~(\ref{eq:14}).\\
\\
\begin{figure}[h]
\includegraphics[width=8.7cm] {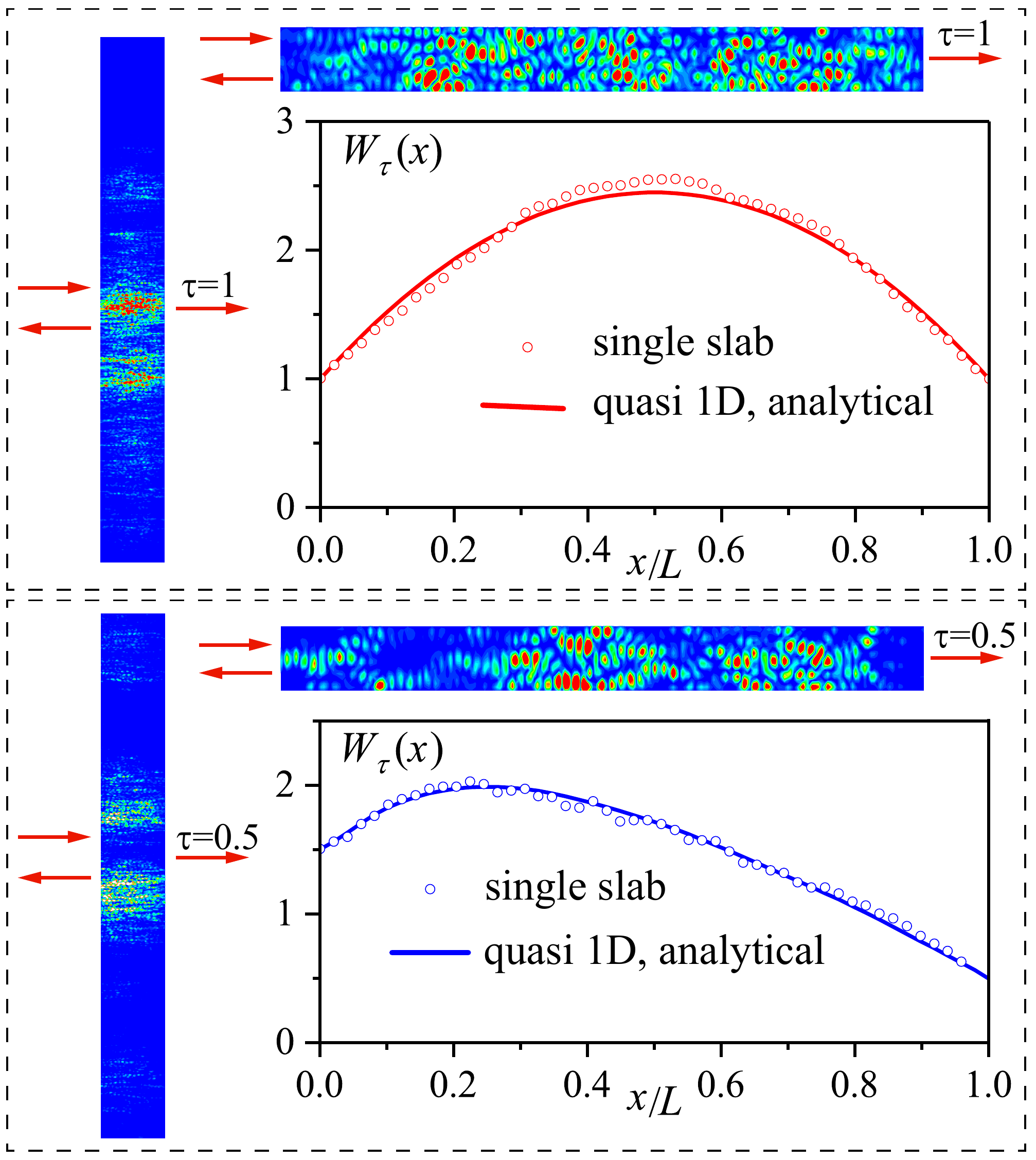}
\caption{Top: Simulations show that the $2$D structure of transparent eigenchannel ($\tau=1$) in a slab (left, the width-to-length ratio $=48$) is very different from that in quasi $1$D (upper right, the ratio $=0.1$): the former exhibits localization in the transverse ($y$) direction while the latter not. Surprisingly, despite this difference, upon integrating $y$ we find that the $1$D structure $W_\tau(x)$ of a single slab -- without ensemble averaging -- is well described by Eq.~(\ref{eq:S3}) for the ensemble of quasi $1$D disorder media (lower right). Bottom: the same as top, but $\tau=0.5$.}
\label{fig:S3}
\end{figure}

\noindent {\bf S2 Eigenchannel structures in a slab}\\

As the impacts of high dimension on eigenchannel structures $\{W_{\tau_n}(x)\}$ remain largely unexplored, in this section we study the explicit forms of $W_{\tau_n}(x)$ in a single disordered slab and their dependence on $N,\omega$. Specifically, we will use the numerical method described in Sec. S7 to simulate the profiles, $\{|E_{n}(x,y)|^2\}$ and $\{W_{\tau_n}(x)\}$, respectively, and compare the results for slabs with those for quasi $1$D media.

The simulation results are shown in Fig.~\ref{fig:S3}. We see that an eigenchannel in a slab and the one in a quasi $1$D medium, even they correspond to the same $\tau$, have totally different $2$D structures: in a slab localization structures appear in the transverse ($y$) direction while in quasi $1$D not. This notwithstanding, surprisingly, when integrate out $y$ we find that the ensuing $1$D structure $W_{\tau_n}(x)$ -- despite it is for a single slab -- is well described by a universal analytic formula $W_\tau(x)$ originally derived for an ensemble of quasi $1$D disordered media \cite{Tian15SM}. For the convenience below we include the formula here. Let the transmission eigenvalue be parametrized as
\begin{equation}\label{eq:S2}
    \tau=\frac{1}{\cosh^2\phi},\quad \phi\geq 0.
\end{equation}
Then the analytic expression of $W_\tau(x)$ is as follows:
\begin{eqnarray}
  W_\tau(x) = S_\tau(x)W_{\tau=1}(x),
  \label{eq:S103}
\end{eqnarray}
where
\begin{eqnarray}
  W_{\tau=1}(x) = 1+\frac{\pi x(L-x)}{2L\ell}
  \label{eq:S104}
\end{eqnarray}
(recall that $\ell$ is the transport mean free path.) and
\begin{eqnarray}
  S_\tau(x) = 2\frac{\cosh^2(h(x')(1-x')\phi)}{\cosh^2(h(x')\phi)}-\tau
  \label{eq:S3}
\end{eqnarray}
with $x'=x/L$. Here $h(x')$ is a function that increases monotonically from $h(1)=1$ as $x'$ decreases from $1$. (For diffusive waves) both $W_{\tau=1}(x)$ and $h(x')$ are independent of $N$.

Because in a single disordered slab the transparent eigenchannel structure is in agreement with Eq.~(\ref{eq:S104}). When fitting the former by the latter, we obtain $\ell$ numerically for a single disorder realization.
\\
\\
\noindent {\bf S3 Lipschitz continuity of $\boldsymbol{W(x;\omega)}$}\\

In this section we will study the Lipschitz continuity of $W(x;\omega)$ (as a function of $\omega$). Specifically, we will show the inequality (\ref{eq:2}) and derive an analytic expression of $\tilde c(x)$. Below to keep the proof simple we will use the result shown numerically in Sec.~S2, i.e., that the analytic expression of $W_\tau(x)$ derived originally for quasi $1$D disordered media \cite{Tian15SM} applies also for a single disordered slab. However, as we will discuss in the end of this section, the key result actually does not depend on the details of the expression, but on its general properties regarding the continuity.

The proof includes following three steps.\\
\\
\noindent{\bf S3.1 Step I}\\

We introduce a family of real-valued functions (labeled by $x$) defined as follows:
\begin{equation}\label{eq:S5}
    g_{z}(x)\equiv W_\tau(x)=W_{z^2}(x),\quad z\equiv\sqrt{\tau}.
\end{equation}
In this step we will study the Lipschitz continuity of this family of functions.

As discussed in Sec.~S1.1, if a function is continuously differentiable, then it is Lipschitz. Thus we consider the derivative of $g_{z}$, which is
\begin{equation}\label{eq:S6}
    \frac{d}{dz}g_{z}(x)=\frac{d}{dz}W_{z^2}(x).
\end{equation}
Substituting Eqs.~(\ref{eq:S103})-(\ref{eq:S3}) into Eq.~(\ref{eq:S6}), we obtain
\begin{eqnarray}
\label{eq:S7}
  \frac{d}{dz}g_{z}(x)=2F(\phi)W_{\tau=1}(x),
\end{eqnarray}
where
\begin{eqnarray}
\label{eq:S8}
  F(\phi)=\frac{\cosh^2\phi}{\sinh\phi}
  \frac{\cosh((1-x')h(x')\phi)}{\cosh^3(h(x')\phi)}h(x')\quad\quad\nonumber\\
  \times (x'\sinh((2-x')h(x')\phi)+(2-x')\sinh(x'h(x')\phi))\nonumber\\
  -\frac{1}{\cosh\phi}.\quad\quad\quad\quad\quad
\end{eqnarray}
For fixed $x'\in [0,1]$ the function $F(\phi)$ is continuous and finite at $\phi=0$. On the other hand, it is well known \cite{Dorokhov84SM,Mello88SM}
that $\phi$ has an upper bound $\phi_m$ which is independent of $N$. Thus the maximal value of $|F(\phi)|$ is independent of $N$. Taking this into account, we find from Eq.~(\ref{eq:S7}) that the derivative of $g_{z}(x)$, with
\begin{equation}\label{eq:S24}
    z\in \mathscr{C}\equiv[1/\cosh\phi_m,1],
\end{equation}
has a least upper bound which is independent of $N$. This bound gives the Lipschitz constant,
\begin{equation}\label{eq:S82}
    \|g_x\|_{\rm Lip}={\rm sup}_{z\in \mathscr{C}}\left|\frac{d}{dz}g_z(x)\right|,
\end{equation}
of the function $g_{z}(x)$. As a result,
\begin{equation}\label{eq:S9}
    |g_{z}(x)-g_{z'}(x)|\leq \|g_x\|_{\rm Lip}|z-z'|,\, \forall\, z,z'\in \mathscr{C}.
\end{equation}
We thus achieve the result of the first step.\\
\\
\noindent{\bf S3.2 Step II}\\

To proceed we introduce the Hilbert-Schmidt (HS) norm of a matrix $A=\{A_{ij}\in \mathds{C}\}$, defined as
\begin{equation}\label{eq:S10}
    \|A\|_{\rm HS}\equiv \sqrt{\sum_{ij}|A_{ij}|^2}=\sqrt{{\rm Tr}(A^\dagger A)}.
\end{equation}
In this step we will bound $|W(x;\omega)-W(x;\omega')|$ by $\|t(\omega)-t(\omega')\|_{\rm HS}$.

By using the Cauchy-Schwartz inequality, we have
\begin{eqnarray}\label{eq:S13}
    &&|W(x;\omega)-W(x;\omega')|\nonumber\\
    &=&\frac{1}{N}\left|\sum_{n=1}^N\left(W_{\tau_n(\omega)}(x)-W_{\tau_n(\omega')}(x)\right)\right|\nonumber\\
    &\leq&\frac{1}{\sqrt{N}}\left(\sum_{n=1}^N\left(W_{\tau_n(\omega)}(x)-W_{\tau_n(\omega')}(x)\right)^2\right)^{1/2}.\quad
\end{eqnarray}
With the HS norm we can rewrite the inequality (\ref{eq:S13}) as
\begin{eqnarray}\label{eq:S14}
    &&|W(x;\omega)-W(x;\omega')|\nonumber\\
    &\leq&\frac{1}{\sqrt{N}}\left\|W_{tt^\dagger(\omega)}(x)-W_{tt^\dagger(\omega')}(x)\right\|_{\rm HS}.
\end{eqnarray}
Throughout this section we put explicitly the dependence of the $N\times N$ matrix $t$ and the eigenvalue spectrum $\{\tau_n\}$ on the disorder realization $\omega$.

To proceed we show the following\\
\\
\noindent{Lemma 3.1.} {\it Let $\mathscr{C}$ be a subset of $\mathds{R}$ and $f:\mathscr{C}\rightarrow \mathds{R}$ be Lipschitz, i.e.,
\begin{equation}\label{eq:S11}
    |f(z)-f(z')|\leq \|f\|_{{\rm Lip}}\,|z-z'|,\quad \forall\, z,z'\in \mathscr{C}.
\end{equation}
Then for any pair $A,A'$ of $N\times N$ hermitian matrices whose eigenvalues are in $\mathscr{C}$, we have}
\begin{equation}\label{eq:S12}
    \|f(A)-f(A')\|_{\rm HS}\leq \|f\|_{{\rm Lip}}\,\|A-A'\|_{\rm HS}.
\end{equation}

{\it Proof.} We follow the scheme described in Ref.~\cite{Haaergrup03SM}.
Because $A$ is hermitian we can write it as
\begin{equation}\label{eq:S15}
    A=\sum_{n=1}^N \lambda_n {\cal P}_n,\quad {\cal P}_n\equiv |n\rangle\langle n|,
\end{equation}
where $\{{\cal P}_n\}$ are $1$D mutually orthogonal projectors, and $|n\rangle$ is the eigenvector corresponding to the eigenvalue $\lambda_n\in \mathscr{C}$. For $A'$ we have the eigenvalue equation,
\begin{equation}\label{eq:S16}
    A'|\psi_n\rangle=\mu_n |\psi_n\rangle.
\end{equation}
In the bases of $\{|n\rangle\}$, the eigenvector $|\psi_n\rangle$ corresponding to $\mu_n\in \mathscr{C}$ can be written as
\begin{equation}\label{eq:S17}
    |\psi_n\rangle=\sum_{n'=1}^N C_{nn'}|n'\rangle, \quad \sum_{n'=1}^N |C_{nn'}|^2=1.
\end{equation}
From this we have the following expression for $A'$,
\begin{equation}\label{eq:S18}
    A'=\sum_{m=1}^N \mu_m {\cal P}'_m,\, {\cal P}'_m\equiv \sum_{n,n'=1}^N C_{mn}C_{mn'}^* |n\rangle\langle n'|.
\end{equation}

By using Eqs.~(\ref{eq:S15}) and (\ref{eq:S18}), we obtain
\begin{eqnarray}
\label{eq:S19}
  && \|f(A)-f(A')\|_{\rm HS}^2\nonumber\\
  &=& {\rm Tr}\bigg(\sum_{m=1}^N \left(f^2(\lambda_m){\cal P}_m+f^2(\mu_m){\cal P}_m'\right)\nonumber\\
  &&-2\sum_{m,m'=1}^N f(\lambda_m)f(\mu_{m'}){\cal P}_{m}{\cal P}_{m'}'\bigg).
\end{eqnarray}
With the insertion of the identity:
\begin{equation}\label{eq:S81}
    \sum_{m=1}^N{\cal P}_{m}=\sum_{m=1}^N{\cal P}_{m}'=1,
\end{equation}
Eq.~(\ref{eq:S19}) is reduced to
\begin{eqnarray}
\label{eq:S20}
  && \|f(A)-f(A')\|_{\rm HS}^2\nonumber\\
  &=& \sum_{m,m'=1}^N \left(f(\lambda_m)-f(\mu_m)\right)^2{\rm Tr}\left({\cal P}_{m}{\cal P}_{m'}'\right).
\end{eqnarray}
On the other hand,
\begin{eqnarray}
\label{eq:S21}
  {\rm Tr}\left({\cal P}_{m}{\cal P}_{m'}'\right)
  &=& {\rm Tr} \sum_{n,n'=1}^N \left(C_{m'n}C_{m'n'}^* |m\rangle\langle m |n\rangle\langle n'|\right)\nonumber\\
  &=& |C_{m'm}|^2\nonumber\\
  &\geq& 0.
\end{eqnarray}
Thus all the summands in Eq.~(\ref{eq:S20}) are nonnegative. With this result and the condition: $\lambda_m,\mu_m\in \mathscr{C}$, we can apply the inequality (\ref{eq:S11}) to Eq.~(\ref{eq:S20}) and obtain
\begin{eqnarray}
\label{eq:S22}
  &&\|f(A)-f(A')\|_{\rm HS}^2\nonumber\\
  &\leq& \|f\|_{\rm Lip}^2
  \sum_{m,m'=1}^N \left(\lambda_m-\mu_m\right)^2{\rm Tr}\left({\cal P}_{m}{\cal P}_{m'}'\right).
\end{eqnarray}
For the sum we can apply the identity (\ref{eq:S20}) with $f$ being the identity map. As a result,
\begin{eqnarray}
\label{eq:S23}
  \sum_{m,m'=1}^N \left(\lambda_m-\mu_m\right)^2{\rm Tr}\left({\cal P}_{m}{\cal P}_{m'}'\right)=\|A-A'\|_{\rm HS}^2.
\end{eqnarray}
Combining it with the inequality (\ref{eq:S22}) we complete the proof of lemma. $\Box$

\noindent {\it Remark.} The lemma differs from a well-known result \cite{Haaergrup03SM}
in the conditions. That is, we require $\mathscr{C}$ to be a subset of $\mathds{R}$, instead of $\mathds{R}$, and correspondingly, the eigenvalues of $A,A'$ to be in this subset. This modification is important for subsequent analysis.

Let $\mathscr{C}$ in Lemma 3.1 be $[1/\cosh\phi_m,1]$ and $f(z)$ be $g_z(x)$. Consider the hermitian matrices, (the derivations below follow Ref.~\cite{Haaergrup03SM}.)
\begin{eqnarray}
\label{eq:S25}
  \tilde t(\omega)=\left(
                     \begin{array}{cc}
                       0 & t^\dagger(\omega) \\
                       t(\omega) & 0 \\
                     \end{array}
                   \right),\,
  \tilde t(\omega')=\left(
                     \begin{array}{cc}
                       0 & t^\dagger(\omega') \\
                       t(\omega') & 0 \\
                     \end{array}
                   \right).\,\,
\end{eqnarray}
It is easy to see that their eigenvalues are $\{\sqrt{\tau_n(\omega)}\}$ and $\{\sqrt{\tau_n(\omega')}\}$, respectively, and thus are in the set of $\mathscr{C}$. Using the inequality (\ref{eq:S9}) and Lemma 3.1 we find that
\begin{equation}\label{eq:S26}
    \|g_{\tilde t(\omega)}(x)-g_{\tilde t(\omega')}(x)\|_{\rm HS}
    \leq \|g_x\|_{\rm Lip}\,\|\tilde t(\omega)-\tilde t(\omega')\|_{\rm HS}.
\end{equation}
By definition (\ref{eq:S5}) we have
\begin{eqnarray}
\label{eq:S27}
  &&g_{\tilde t(\omega)}-g_{\tilde t(\omega')} \nonumber\\
  &=& W_{\tilde t^2(\omega)}-W_{\tilde t^2(\omega')} \nonumber\\
  &=& \left(
        \begin{array}{cc}
          W_{t^\dagger t(\omega)}-W_{t^\dagger t(\omega')} & 0 \\
          0 & W_{tt^\dagger(\omega)}-W_{tt^\dagger(\omega')} \\
        \end{array}
      \right),\,\,\,
\end{eqnarray}
where to make the formula compact we have suppressed the parameter $x$. Substituting it into the inequality (\ref{eq:S26}) we obtain
\begin{eqnarray}
\label{eq:S28}
  \|W_{t^\dagger t(\omega)}-W_{t^\dagger t(\omega')}\|_{\rm HS}^2+\|W_{tt^\dagger(\omega)}-W_{tt^\dagger(\omega')}\|_{\rm HS}^2\quad\nonumber\\
  \leq \|g_x\|_{\rm Lip}^2\left(\|t(\omega)-t(\omega')\|_{\rm HS}^2+\|t^\dagger(\omega)-t^\dagger(\omega')\|_{\rm HS}^2\right),\,\,
\end{eqnarray}
from which
\begin{eqnarray}
\label{eq:S29}
  &&\|W_{tt^\dagger(\omega)}(x)-W_{tt^\dagger(\omega')}(x)\|_{\rm HS}\nonumber\\
  &\leq& \|g_x\|_{\rm Lip}\|t(\omega)-t(\omega')\|_{\rm HS}
\end{eqnarray}
follows.

Combining the inequalities (\ref{eq:S14}) and (\ref{eq:S29}), we obtain
\begin{eqnarray}\label{eq:S30}
|W(x;\omega)-W(x;\omega')|
\leq\frac{\|g_x\|_{\rm Lip}}{\sqrt{N}}\left\|t(\omega)-t(\omega')\right\|_{\rm HS}.
\end{eqnarray}
We thus achieve the result of the second step.\\
\\
\noindent {\bf S3.3 Step III}\\

In the last step we will further bound $\left\|t(\omega)-t(\omega')\right\|_{\rm HS}$ on the right-hand side of the inequality (\ref{eq:S30}) by $\|\omega-\omega'\|$. To this end we note that the matrix element of $t(\omega)$ is given by
\begin{eqnarray}
\label{eq:S31}
  t_{ab}(\omega)
  &=& -i\sqrt{\tilde v_a\tilde v_b} \int\!\!\!\!\int dydy' \varphi_a(y)\varphi_b^*(y')\nonumber\\
  &\times& G(x=\infty,y,x'=-\infty,y')\nonumber\\
  &\equiv&-i\sqrt{\tilde v_a\tilde v_b}\langle \infty a|G(\omega)|-\infty b\rangle,
\end{eqnarray}
where in the last line we have put the $\omega$-dependence of Green's function $G$ explicitly. With the substitution of this expression, $\left\|t(\omega)-t(\omega')\right\|_{\rm HS}$ is written as
\begin{eqnarray}
\label{eq:S32}
  && \left\|t(\omega)-t(\omega')\right\|_{\rm HS}^2\nonumber\\
  &=& \sum_{a,b=1}^N \tilde v_a\tilde v_b |\langle \infty a|G(\omega)-G(\omega')|-\infty b\rangle|^2.
\end{eqnarray}
With the help of the identity,
\begin{eqnarray}\label{eq:S33}
G(\omega)-G(\omega')
&=&G(\omega)\left(G^{-1}(\omega')-G^{-1}(\omega)\right)G(\omega')\quad\nonumber\\
&=&G(\omega)\left(\omega-\omega'\right)G(\omega'),
\end{eqnarray}
we write Eq.~(\ref{eq:S32}) as
\begin{eqnarray}
\label{eq:S34}
  && \left\|t(\omega)-t(\omega')\right\|_{\rm HS}^2\nonumber\\
  &=& \sum_{a,b=1}^N \tilde v_a\tilde v_b |\langle \infty a|G(\omega)\left(\omega-\omega'\right)G(\omega')|-\infty b\rangle|^2.\quad\quad
\end{eqnarray}
After simple algebra we further reduce it to
\begin{equation}\label{eq:S35}
    \left\|t(\omega)-t(\omega')\right\|_{\rm HS}^2=(\omega-\omega')^{\rm T}{\cal M}(\omega,\omega')(\omega-\omega').
\end{equation}
Here ${\cal M}(\omega,\omega')$ is an $M\times M$ (recall that $M$ is the number of lattice points when the disordered medium is discretized.) hermitian matrix. The corresponding matrix element ${\cal M}_{rr'}(\omega,\omega')$, with $r\equiv (x,y)$ and $r'\equiv (x',y')$ being the discrete lattice points, reads
\begin{eqnarray}\label{eq:S36}
    &&{\cal M}_{rr'}(\omega,\omega')\nonumber\\
    &=&\left(G^\dagger(\omega){\cal V}_LG(\omega)\right)_{r'r}\left(G(\omega'){\cal V}_0G^\dagger(\omega')\right)_{rr'},
\end{eqnarray}
where
\begin{eqnarray}
\label{eq:S102}
  {\cal V}_{0}&=&\sum_{a=1}^N \tilde v_a|-\infty a\rangle\langle -\infty a|,\nonumber\\
  {\cal V}_{L}&=&\sum_{a=1}^N \tilde v_a|\infty a\rangle\langle \infty a|.
\end{eqnarray}
Let $\lambda_1(\omega,\omega')$ be the largest eigenvalue of ${\cal M}(\omega,\omega')$ (which obviously is nonnegative). Then by using Eq.~(\ref{eq:S35}) the following inequality,
\begin{equation}\label{eq:S37}
    \left\|t(\omega)-t(\omega')\right\|_{\rm HS}\leq \sqrt{\lambda_1(\omega,\omega')}\|\omega-\omega'\|,
\end{equation}
can be readily shown.

Because $M_{rr'}$ decays rapidly as the distance between $r$ and $r'$ increases, we have
\begin{eqnarray}
{\rm Tr}{\cal M}^n={\cal O}(M),\quad \forall n\in \mathds{N}.
\label{eq:S42}
\end{eqnarray}
Therefore, all the absolute values of the eigenvalues cannot grow with $M$. Thanks to $M\propto N$ we have
\begin{equation}\label{eq:S43}
    \lambda_1(\omega,\omega')={\cal O}(1),
\end{equation}
from which
\begin{equation}\label{eq:S44}
    \tilde \lambda_1\equiv {\rm sup}_{\omega,\omega'}\lambda_1(\omega,\omega')={\cal O}(1)
\end{equation}
follows.

Taking the inequality (\ref{eq:S37}) and the definition of $\tilde \lambda_1$ [cf.~Eq.~(\ref{eq:S44})] into account we obtain
\begin{equation}\label{eq:S39}
    \left\|t(\omega)-t(\omega')\right\|_{\rm HS}\leq \tilde\lambda_1^{1/2}\|\omega-\omega'\|.
\end{equation}
Combining it with the inequality (\ref{eq:S30}) we obtain
\begin{eqnarray}\label{eq:S40}
|W(x;\omega)-W(x;\omega')|
\leq \|W_x\|_{\rm Lip}\|\omega-\omega'\|
\end{eqnarray}
and the analytic expression,
\begin{equation}\label{eq:S41}
    \|W_x\|_{\rm Lip}=\frac{\tilde c(x)}{\sqrt{N}},\quad \tilde c(x)\equiv \tilde\lambda_1^{1/2} \|g_x\|_{\rm Lip}.
\end{equation}
Equations (\ref{eq:S44}) and (\ref{eq:S41}) show that
\begin{equation}\label{eq:S83}
    \|W_x\|_{\rm Lip}={\cal O}(1/\sqrt{N})
\end{equation}
and its depth profile is governed by the eigenchannel structure $W_\tau(x)$. Thus the inequality (\ref{eq:2}) is justified. In the next section we further provide the numerical confirmation of the scaling (\ref{eq:S83}).\\
\\
\noindent{\bf S3.4 Discussions}\\

We have seen that the exact expression of $W_\tau(x)$ is not essential to the proof. What we really used in the proof are the following two properties. (i) The structures $\{W_{\tau_n}(x)\}$ have a universal expression $W_\tau(x)$, which is a function of $\tau$; (ii) $\frac{d}{dz}W_{z^2}(x)$ is bounded. Provided (i) and (ii) are satisfied (according to previous experiences \cite{Tian15SM} and the results in Sec.~S2, this seems to be the case.), even though the exact expression of $W_\tau(x)$ is unknown, we can still repeat the proof and obtain the key results, i.e., the inequality (\ref{eq:2}) and the scaling (\ref{eq:S83}). Of course we cannot determine the explicit form of $\|g_x\|_{\rm Lip}$ and $\tilde c(x)$ in this case.
\\
\\
\begin{figure}[h]
\includegraphics[width=8.7cm] {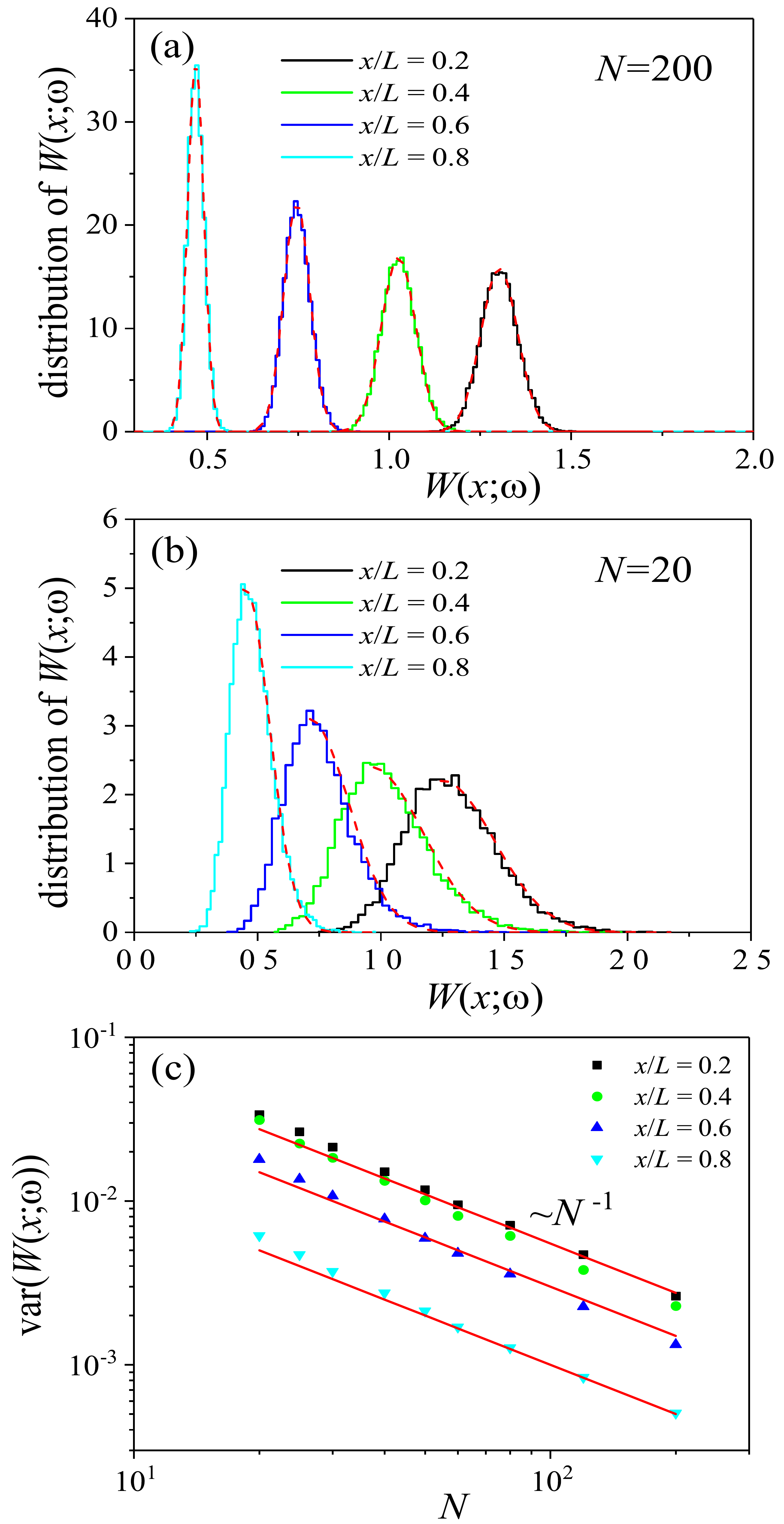}
\caption{We simulate $W(x;\omega)$ for $10^4$ randomly chosen disorder realizations $\omega$. For both large (a) and small (b) $N$ the distributions of $W(x;\omega)$ at distinct $x$ (histograms) are well fit by (one-sided) Gaussian distributions (dashed lines). The variance scales with $N$ as $N^{-1}$. $L=50$}
\label{fig:S2}
\end{figure}
\begin{figure}[h]
\includegraphics[width=8.7cm] {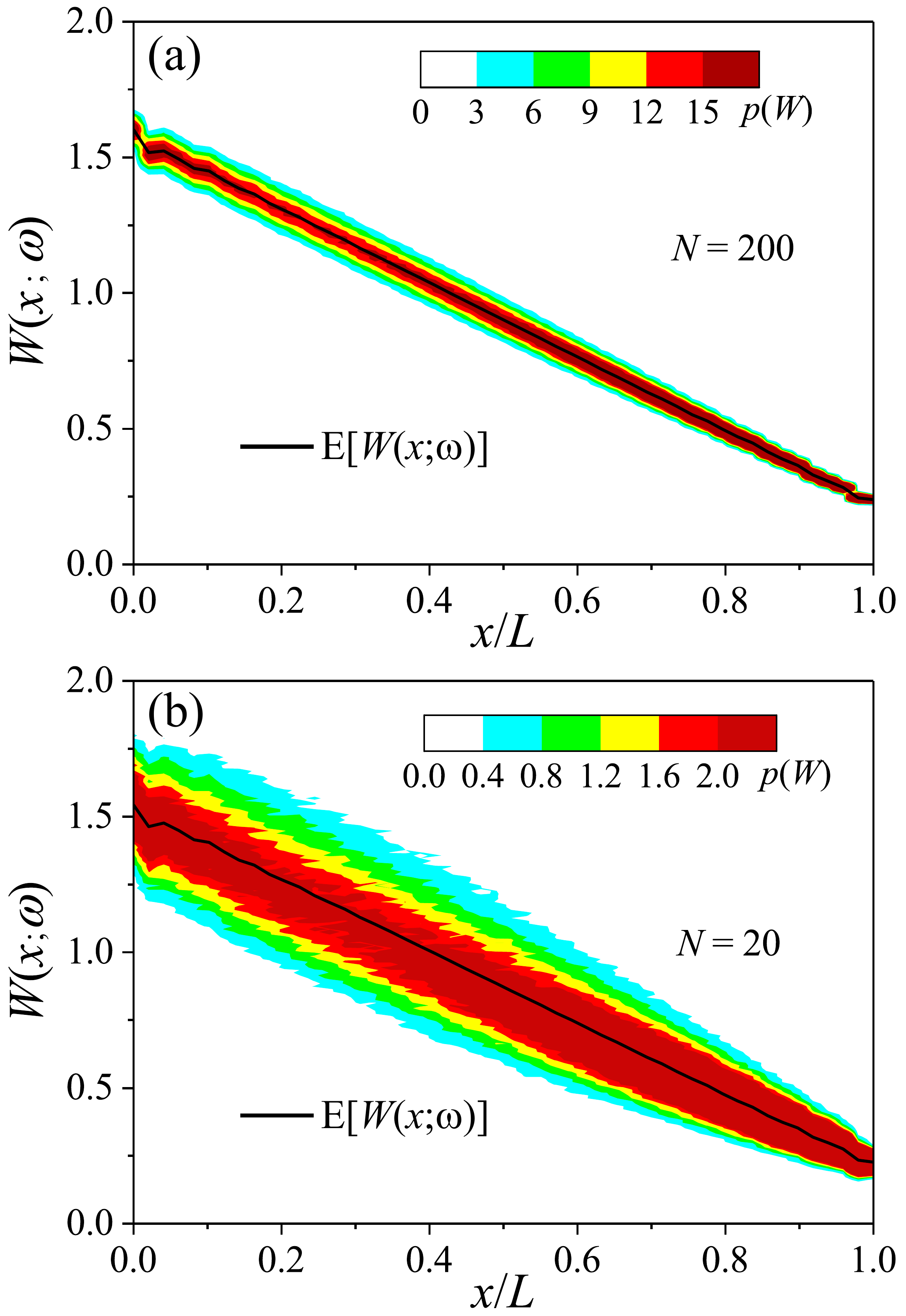}
\caption{The same as Fig.~\ref{fig:S2}. The simulated profiles of the steady state $W(x;\omega)$ concentrate strongly around their disorder mean ${\rm E}[W(x;\omega)]$.}
\label{fig:S5}
\end{figure}
\begin{figure}
\includegraphics[width=8.7cm] {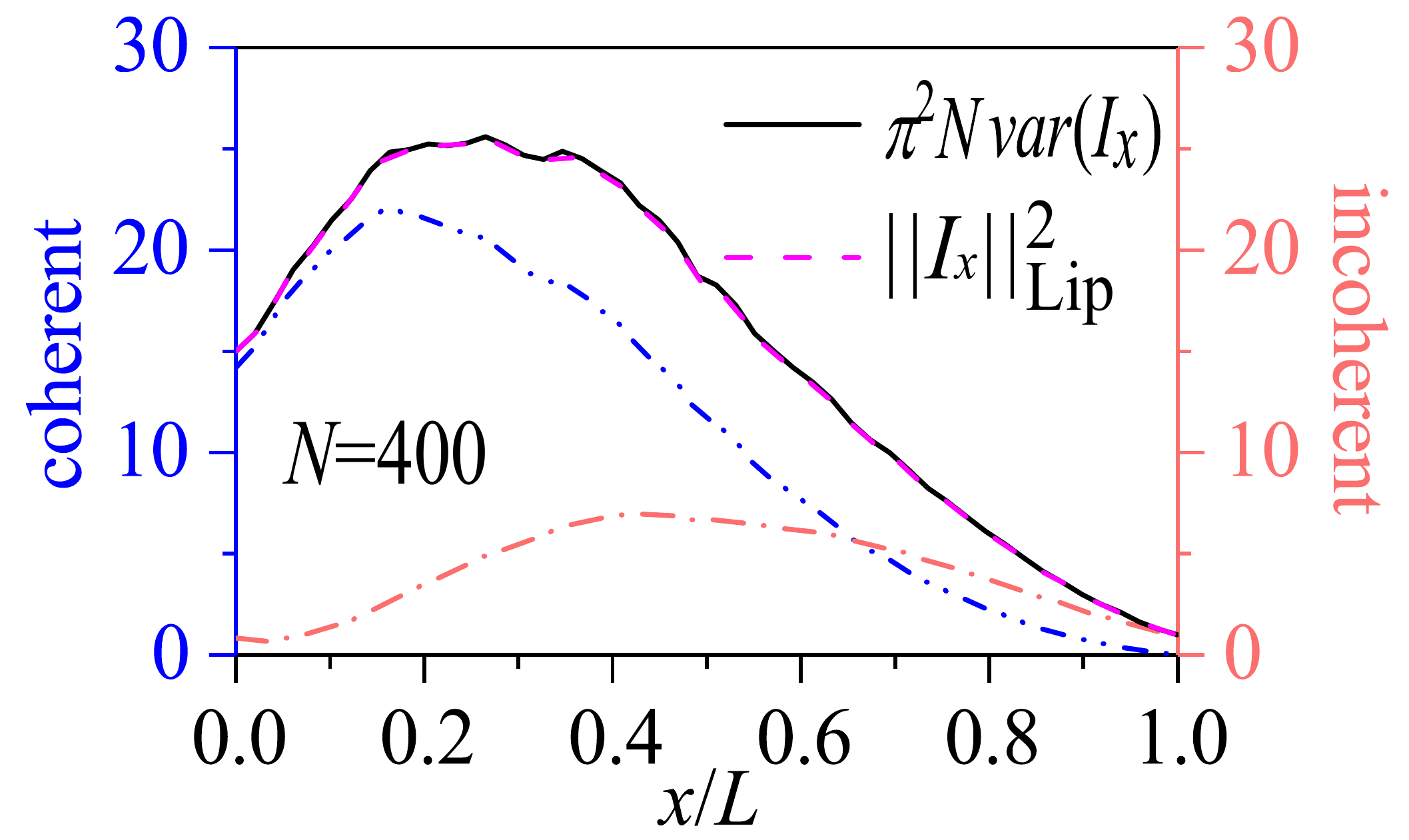}
\caption{The depth profiles of $\pi^2Nvar(I_x)$ obtained from simulating $10^4$ randomly chosen incoming waves and $\|I_x\|_{\rm Lip}^2$ obtained from the analytic formula (\ref{eq:14}) collapse into a single curve. The blue (red) curve is the (in)coherent part of $\|I_x\|_{\rm Lip}^2$.}
\label{fig:S4}
\end{figure}

\noindent{{\bf S4 Proof and numerical confirmation of the concentration inequality (\ref{eq:5})}}\\

\noindent{{\bf S4.1 Proof of the inequality}}\\

{\it Proof.} We use (Pisier's) Theorem 0.4. Let $\varpi$ in that theorem be $\omega$, $f(\varpi)$ be $W(\omega)$, and $k=M$. Due to the inequality (\ref{eq:2}) and recalling that the disorder ensemble is described by a Gaussian probability measure on $\mathds{R}^M$ of zero mean and variance $\sigma^2$, we find that the conditions of Pisier's theorem are satisfied. After rescaling we obtain from the theorem the inequality (\ref{eq:5}). $\Box$\\
\\
\noindent{{\bf S4.2 Numerical confirmation}}\\

We simulate the eigenchannel structures $\{W_{\tau_n}(x)\}$ by using the method described in Sec.~S7 for a large number of randomly chosen $\omega$. For each $\omega$ we substitute the simulation results of $\{W_{\tau_n}(x)\}$ into the expression:
\begin{eqnarray}
W(x;\omega)=\frac{1}{N}\sum_{n=1}^N W_{\tau_n}(x)
\label{eq:S84}
\end{eqnarray}
and obtain $W(x;\omega)$. As shown in Fig.~\ref{fig:S2}, for both large (a) and small (b) $N$ the distributions of $W(x;\omega)$ at distinct $x$ are all well fit by (one-sided) Gaussian distributions. Moreover, the variance of these distributions are found to change with $N$ as (c):
\begin{equation}\label{eq:S85}
    \ln var(W(x;\omega))=-\ln N+\tilde C(x),
\end{equation}
where $\tilde C(x)$ is some function independent of $N$. Thus simulations confirm the scaling (\ref{eq:S83}) and the concentration inequality (\ref{eq:5}). In addition, they confirm that the scaling and the inequality hold for both large and small $N$, as long as the medium is diffusive. As shown in Fig.~\ref{fig:S5}, simulations confirm that for both large and small $N$, the profiles of the steady state $W(x;\omega)$ concentrate strongly around their disorder mean ${\rm E}[W(x;\omega)]$ which is a linear decrease, namely, the diffusive steady state.\\
\\
\noindent {\bf S5 Derivations and numerical studies of Eq.~(\ref{eq:20})}\\

In the first part of this section we derive Eq.~(\ref{eq:20}). For this purpose we decompose $I_x(c)$ into two parts,
\begin{eqnarray}
    I_x(c)&=& I_{inc,x}(c)+I_{coh,x}(c),\label{eq:S70}\\
    I_{inc,x}(c)&\equiv&\sum_{n=1}^N |c_n|^2 W_{\tau_n}(x),\label{eq:S71}\\
    I_{coh,x}(c)&\equiv&\sum_{n\neq n'}^N c_n^*c_{n'} E_{n}^\dagger(x)\cdot E_{n'}(x).
    \label{eq:S72}
\end{eqnarray}
The first part, $I_{inc,x}(c)$, accounts for the incoherent contributions of eigenchannels, while the second part, $I_{coh,x}(c)$, for the contributions arising from the coherence between distinct eigenchannels. By using the integral (\ref{eq:S79}) in Sec.~S8, we find
\begin{eqnarray}
\label{eq:S73}
  var(I_x) = \int \left(I_x(c)-W(x;\omega)\right)^2 d\mu\quad\quad\quad\nonumber\\
  = \int \left(\left(I_{inc,x}(c)-W(x;\omega)\right)+I_{coh,x}(c)\right)^2 d\mu\quad\quad\nonumber\\
  = \int \left(I_{inc,x}(c)-W(x;\omega)\right)^2 d\mu+\int I_{coh,x}^2(c) d\mu.\,
\end{eqnarray}
This shows that the variance includes the incoherent and coherent contributions, corresponding to the first and second terms, respectively. Substituting Eqs.~(\ref{eq:S71}) and (\ref{eq:S72}) into these two terms, and using the integrals (\ref{eq:S79})-(\ref{eq:S61}), we obtain
\begin{eqnarray}
\label{eq:S74}
  &&\int \left(I_{inc,x}(c)-W(x;\omega)\right)^2 d\mu\nonumber\\
  &=& \frac{1}{N(N+1)}\sum_{n=1}^N\left(W_{\tau_n}(x)-W(x;\omega)\right)^2
\end{eqnarray}
and
\begin{eqnarray}
\label{eq:S75}
  \int I_{coh,x}^2(c)d\mu=\frac{1}{N(N+1)}\sum_{n\neq n'}^N|E_{n}^\dagger(x)\cdot E_{n'}(x)|^2.
\end{eqnarray}
Adding Eqs.~(\ref{eq:S74}) and (\ref{eq:S75}) together we find
\begin{eqnarray}
\label{eq:S93}
  var(I_x) &=& \frac{1}{N(N+1)}\bigg(\sum_{n=1}^N\left(W_{\tau_n}(x)-W(x;\omega)\right)^2\nonumber\\
  &&\quad\quad\quad\,\,\,\,\,\,\,\,+\sum_{n\neq n'}^N|E_{n}^\dagger(x)\cdot E_{n'}(x)|^2\bigg).
\end{eqnarray}
Comparing it with Eq.~(\ref{eq:14}) we justify Eq.~(\ref{eq:20}).

We have provided in Fig.~\ref{fig:1} simulation results for distinct $N$ and $x$, which confirm the relation (\ref{eq:20}). In the second part of this section we perform further numerical studies of $var(I_x)$ and $\|I_x\|_{\rm Lip}^2$. To be specific, we fix $N$ to be $400$, and use the method described in Sec.~S7 to simulate wave propagation in a single disordered slab ($L=50$) for $10^4$ randomly chosen incoming waves. The simulated depth profile of $\pi^2Nvar(I_x)$ is shown in Fig.~\ref{fig:S4}. Furthermore, for the same disordered medium we use the analytic formula (\ref{eq:14}) to calculate $\|I_x\|_{\rm Lip}^2$, and its incoherent and coherent parts, respectively. Corresponding profiles are shown in Fig.~\ref{fig:S4} also. We see that the profiles of $\pi^2Nvar(I_x)$ and $\|I_x\|_{\rm Lip}^2$ are identical, and the incoherent (coherent) part of $\|I_x\|_{\rm Lip}^2$ dominates in the back (front) part of the medium.\\
\\
\noindent {\bf S6 Derivations of Eq.~(\ref{eq:13})}\\

In most of this work we have focused on the energy density $I_x(c)$ integrated over the transverse coordinate $y$. In this section we consider a generic observable, namely, an operator $\hat O$. It is important that this observable needs not to be an integral over $y$. For simplicity we consider $\hat O$ which is hermitian. Note that this condition is not essential. For a non-hermitian operator its expectation value at a stationary scattering state has an imaginary part in general. In this case we only need to discuss the real and imaginary parts separately, and repeat the analysis below.

Treating $\omega$ as a scattering potential, we apply the scattering theory of waves \cite{Newton82SM} to Eq.~(\ref{eq:15}) and find
\begin{equation}\label{eq:S122}
    \hat v_x^{\frac{1}{2}}E(x,y)=\sum_{a=1}^N(t(x)c)_a\varphi^*_a(y).
\end{equation}
Recall that
\begin{equation}\label{eq:S123}
    c=\sum_{n=1}^N c_n v_n
\end{equation}
is the incoming current amplitude in the eigenchannel representation, and $\hat v_x\equiv (1-\partial_{\Omega y}^2)^{\frac{1}{2}}$ is a scalar (not vector) operator accounting for the group velocity in the waveguide modes. Using Eqs.~(\ref{eq:S122}) and (\ref{eq:S123}), we find that at the stationary scattering state $E(x,y)$ corresponding to $c$, the expectation value of $\hat O$ is
\begin{eqnarray}\label{eq:S124}
  O(c)\equiv \langle E|\hat O|E\rangle=\sum_{n,n'=1}^N c_n^*c_{n'} \langle E_n|\hat O|E_{n'}\rangle.
\end{eqnarray}
Recall that $E_n(x,y)$ is the $2$D wave field of the $n$th eigenchannel. Strictly speaking, like the definition of $I_x(c)$, in the last equality above there are two scalar operators $\hat v_x^{-\frac{1}{2}}$ sandwiching $\hat O$. However, because these two operators account only for an irrelevant overall factor, we omit them hereafter. Naturally, the problem now is: {\it For fixed $\omega$, does $O(c)$ exhibit universal behaviors when $c$ varies}? Below we use L$\acute{\rm e}$vy's lemma to study this problem.

Equation (\ref{eq:S124}) defines a real-valued function over $S^{2N-1}$. If the coefficients $\langle E_n|\hat O|E_{n'}\rangle$ do not diverge, (this condition can be achieved readily in physical systems.) then $O(c)$ is continuously differentiable and thus Lipschitz, according to the discussions in Sec.~S1.1. With the help of (L$\acute{\rm e}$vy) Lemma 0.5, we obtain the following concentration inequality:
\begin{equation}\label{eq:S125}
    {\rm Pr}\left(\left|O(c)-{\bar O}\right|>\varepsilon\right)\leq 2e^{-
\frac{\delta\varepsilon^2 N}{\|O\|_{{\rm Lip}}^2}},
\end{equation}
where $\|O\|_{{\rm Lip}}$ is the Lipschitz constant of $O$. This means that $O(c)$ concentrates around its mean: ${\bar O}\equiv\int Od\mu$. The latter can be readily calculated by using the definition (\ref{eq:S124}). As a result,
\begin{equation}\label{eq:S126}
    {\bar O}=\frac{1}{N}\sum_{n=1}^N \langle E_n|\hat O|E_n\rangle.
\end{equation}
The wave-to-wave fluctuations of $O$ is governed by $\|O\|_{{\rm Lip}}$. According to the inequality (\ref{eq:S126}), provided
\begin{equation}\label{eq:S127}
    {\bar O}\gg\frac{\|O\|_{{\rm Lip}}}{\sqrt{N}},
\end{equation}
the concentration effect is strong and the wave-to-wave fluctuations are negligible. Thus we have Eq.~(\ref{eq:13}). Note that the condition (\ref{eq:S127}) is the generalization of the criterion (\ref{eq:7}).\\
\\
\noindent {\bf S7 Method of numerical simulations}\\

We simulate the wave propagation by using Eq.~(\ref{eq:15}), where $\delta \epsilon (x,y)$ is drawn from a uniform distribution over an interval $[-\delta \epsilon_0,\delta \epsilon_0]$. Here $\delta \epsilon_0\in (0,1)$ governs the disorder strength and is set to $0.97$ in simulations. The values of $\delta \epsilon (x,y)$ at distinct spatial points are chosen in the same way and independently. For simulations, Eq.~(\ref{eq:15}) is discretized on a square grid, with the grid spacing being the inverse wave number in the ideal waveguide, and $L$ is defined as the number of discrete points in $x$-direction. We solve this equation by using the recursive Green's function method \cite{MacKinnon85SM}. We first calculate $\langle x=L,y|G|x'=0,y'\rangle$, from which we obtain the transmission matrix $t$. By performing the singular value decomposition of $t$, we obtain $\{v_n\}$ and $\{\tau_n\}$. Next, we calculate $\langle x,y|G|x'=0,y'\rangle$, from which we obtain the matrix $t(x)$. Then the wave field in the interior of the medium is given by $E(x,y)=\sum_{a=1}^N (t(x)c)_a\varphi^*_a(y)$. For a single disorder realization, small oscillations in $x$ are often superposed on non-oscillatory backgrounds. These oscillations occur in the wavelength scale and are unimportant. They are removed by performing the local (in $x$) average over a window with a width of wavelength.

To draw a point $c\equiv(c_1,c_2,\cdots,c_N)$ from a uniform distribution over $ S^{2N-1}$, we use the standard method \cite{Milman86SM,Marsaglia72SM} (cf.~Sec.~S0.1).
First, we generate $2N$ independent standard normal random variables, $(a'_n,b'_n)$ with $n=1,2,\cdots, N$. Secondly, we normalize each $a'_n$ ($b'_n$) by $\sqrt{\sum_{n=1}^N((a'_n)^2+(b'_n)^2)}$. Define the normalized $a'_n$ ($b'_n$) as the real (imaginary) part of $c_n$ [cf.~Eq.~(\ref{eq:S58})]. We generate a desired random point $c$.\\
\\
\noindent{\bf S8 Derivations of three integrals}\\

In the discussions above we have used the following three integrals:
\begin{eqnarray}
  \int c_nc_{n'}^*d\mu &=& \frac{1}{N}\delta_{nn'},\label{eq:S79}\\
  \int |c_n|^4d\mu &=& \frac{2}{N(N+1)},\label{eq:S60}\\
  \int |c_n|^2|c_{n'}|^2d\mu &=& \frac{1}{N(N+1)},\quad n\neq n'.
  \label{eq:S61}
\end{eqnarray}
Recall that $\mu$ is the uniform probability measure over $S^{2N-1}$, and $(c_1,c_2,\cdots,c_N)$ are the coordinates of $S^{2N-1}$. In this section we derive these integrals.

The first integral is easy to show. For $n\neq n'$ the left-hand side of Eq.~(\ref{eq:S79}) vanishes after integrating out the relative phase between $c_n$ and $c_{n'}$. Thus $n,n'$ must be paired, i.e., $n=n'$ for it not to vanish. By the symmetry Eq.~(\ref{eq:S79}) then follows. The remainder of this section is to derive the last two integrals.

We parametrize $c_n$ by [cf.~Eq.~(\ref{eq:S58})]:
\begin{eqnarray}
    &&c_n=a_n+ib_n,\quad a_n,b_n\in \mathds{R}\nonumber\\
    &&\sum_{n=1}^N \left(a_n^2+b_n^2\right)=1,
    \label{eq:S62}
\end{eqnarray}
where $(a_1,b_2,\cdots,a_N,b_N)$ constitute the real coordinates of $S^{2N-1}$. Substituting it into the left-hand sides of Eqs.~(\ref{eq:S60}) and (\ref{eq:S61}) and using the symmetry, we obtain
\begin{eqnarray}
\int |c_n|^4 d\mu &=& 2\left(\int a_1^4d\mu+\int a_1^2b_1^2 d\mu\right),\label{eq:S98}\\
\int |c_n|^2|c_{n'}|^2d\mu &=&4\int a_1^2b_1^2d\mu.
  \label{eq:S99}
\end{eqnarray}
So the calculations of the two integrals are reduced to the calculations of the integrals $\int a_1^4d\mu$ and $\int a_1^2b_1^2 d\mu$.

To calculate the last two integrals we define
\begin{equation}\label{eq:S100}
    r_{2n-1}\equiv a_n,\quad r_{2n}\equiv b_n,\quad n=1,2,\cdots,N
\end{equation}
and parametrize $(r_1,r_2,\cdots,r_{2N-1},r_{2N})$ as follows \cite{Prudnikov81}:
\begin{eqnarray}
r_1&=&\cos\varphi_1,\nonumber\\
r_2&=&\sin\varphi_1\cos\varphi_2,\nonumber\\
r_3&=&\sin\varphi_1\sin\varphi_2\cos\varphi_3,\nonumber\\
&&\cdots\cdots\cdots\cdots\cdots\nonumber\\
r_{2N-1}&=&\sin\varphi_1\sin\varphi_2\cdots\sin\varphi_{2N-2}\cos\varphi_{2N-1},\nonumber\\
r_{2N}&=&\sin\varphi_1\sin\varphi_2\cdots\sin\varphi_{2N-2}\sin\varphi_{2N-1},
\label{eq:S63}
\end{eqnarray}
where
\begin{equation}\label{eq:S64}
    0\leq \varphi_1,\varphi_2,\cdots,\varphi_{2N-2}\leq \pi,\quad 0\leq \varphi_{2N-1}<2\pi.
\end{equation}
Correspondingly, the Jacobian
\begin{equation}\label{eq:S65}
    J=\sin^{2N-2}\varphi_1\sin^{2N-3}\varphi_2\cdots \sin^{2}\varphi_{2N-3}\sin\varphi_{2N-2}.
\end{equation}
With the substitution of this parametrization we obtain
\begin{eqnarray}
  \int a_1^4d\mu &=& \int r_1^4d\mu = \frac{\int_0^{\frac{\pi}{2}}\cos^4\varphi_1\sin^{2N-2}\varphi_1 d\varphi_1}{\int_0^{\frac{\pi}{2}}\sin^{2N-2}\varphi_1 d\varphi_1}\nonumber\\
  &=& \frac{B\left(\frac{2N-1}{2},\frac{5}{2}\right)}{B\left(\frac{2N-1}{2},\frac{1}{2}\right)},
  \label{eq:S66}
\end{eqnarray}
where $B(x,y)$ is the beta function. Using the identity:
\begin{equation}\label{eq:S67}
    B(x,y)=\frac{\Gamma(x)\Gamma(y)}{\Gamma(x+y)},
\end{equation}
with $\Gamma(x)$ being the gamma function, we reduce Eq.~(\ref{eq:S66}) to
\begin{eqnarray}
  \int a_1^4d\mu=\frac{3}{4N(N+1)}.
  \label{eq:S94}
\end{eqnarray}
With the substitution of this parametrization we also obtain
\begin{eqnarray}
  &&\int a_1^2b_1^2d\mu=\int r_1^2r_2^2d\mu \nonumber\\
  &=& \frac{\int_0^{\pi}\cos^2\varphi_1\sin^{2N}\varphi_1 d\varphi_1}{\int_0^{\pi}\sin^{2N-2}\varphi_1 d\varphi_1}
  \frac{\int_0^{\pi}\cos^2\varphi_2\sin^{2N-3}\varphi_2 d\varphi_2}{\int_0^{\pi}\sin^{2N-3}\varphi_2 d\varphi_2}\nonumber\\
  &=& \frac{B\left(\frac{2N+1}{2},\frac{3}{2}\right)}{B\left(\frac{2N-1}{2},\frac{1}{2}\right)}
  \frac{B\left(\frac{2N-2}{2},\frac{3}{2}\right)}{B\left(\frac{2N-2}{2},\frac{1}{2}\right)}.
  \label{eq:68}
\end{eqnarray}
Using the identity (\ref{eq:S67}) we reduce it to
\begin{eqnarray}
\int a_1^2b_1^2d\mu &=&\frac{1}{4N(N+1)}.
\label{eq:S95}
\end{eqnarray}
Substituting Eqs.~(\ref{eq:S94}) and (\ref{eq:S95}) into Eqs.~(\ref{eq:S98}) and (\ref{eq:S99}) we justify Eqs.~(\ref{eq:S60}) and (\ref{eq:S61}).


\clearpage

\begin{thebibliography}{}
\bibitem{note3} Throughout this work we require the sample size be much larger than the transport mean free path $\ell$. The latter is usually defined for a disorder ensemble. As we show in this work, $\ell$ is well defined for a single disordered medium also, and is identical to that defined for an ensemble. In SM (S2) we further show that, when $N$ is large, for a single disordered medium $\ell$ can be also defined by the transparent eigenchannel structure $W_{\tau=1}(x)$.
\bibitem{note5} The group velocity in the air background is unity. In simulations all lengths are in unit of the inverse wave number in the air background.
\bibitem{note6} Here we write $G$ in the representation which is a mixture of the real coordinate $x$ and the waveguide mode. In other words, we perform the Fourier transform of $\langle xy|G|x'y'\rangle$ with respect to $y,y'$.
\bibitem{note7} To make formulae compact we suppress $\omega$ and $\Omega$ in $I_x(c)$ and $E_n$ throughout.
\bibitem{note2} Strictly speaking, the definition of energy density excludes the scalar operator $\hat v_x$. However, we have found that omitting $\hat v_x$ in the definition only leads to the appearance of an irrelevant overall factor.
\bibitem{note1} For the sample-to-sample fluctuations of $I_{x}(c)$ (with the incoming current amplitude $c$ fixed), we are not aware of any reports of either the exponential distribution of diffusive waves (except for the special case of $x=L$ and $c$ chosen to correpond to a plane wave or a Gaussian beam \cite{Rossum95}) or the stretched-exponential distribution of localized waves.
\end{thebibliography}
\end{document}